\documentstyle[12pt]{article}

\def\theequation{\arabic{section}.\arabic{equation}}

\def\be{\begin{equation}}
\def\ee{\end{equation}}
\def\c{\cite}

\def\={\!=\!}
\def\+{\!+\!}
\def\-{\!-\!}
\def\a{\alpha}
\def\b{\beta}
\def\d{\delta}
\def\s{\sigma}
\def\r{\rho}
\def\p{\partial}
\def\t{\theta}

\begin{document}

%%%%%%%%%%%%%%%%%%%%%%%%

\begin{titlepage}

%%%%%%%%%%%%%%%%%%%%%%%%

\begin{flushright}
\large UAB-FT-359
\end{flushright}

\baselineskip 7ex
\mbox{}\vspace*{1.5ex}

\begin{center}
{\bf \Huge Back reaction in the formation of a straight cosmic string}
\end{center}

\vspace{5ex}

\baselineskip 3.5ex

\begin{center}
{\bf Antonio Campos}\footnote{Institut de F\'{\i}sica d'Altes Energies
(IFAE)} \\ Grup de F\'{\i}sica Te\`orica, \\
Universitat Aut\`onoma de Barcelona, \\
08193 Bellaterra (Barcelona), Spain
\end{center}

\setcounter{footnote}{0}

\begin{center}
{\bf Rosario Mart\'{\i}n} and {\bf Enric Verdaguer}\footnotemark  \\
Departament de F\'{\i}sica Fonamental, \\
Universitat de Barcelona, \\
Av. Diagonal 647, \\
08028 Barcelona, Spain
\end{center}

\vspace{7ex}

\baselineskip 1ex

\begin{abstract}
\noindent A simple model for the formation of a straight cosmic string,
wiggly or unperturbed is considered. The gravitational field of such
string is computed in the linear approximation. The vacuum expectation
value of the stress tensor of a massless scalar quantum field coupled
to the string gravitational field is computed to the one loop order.
Finally, the back-reaction effect on the gravitational field of the
string is obtained by solving perturbatively the semiclassical
Einstein's equations.
\end{abstract}
\vspace{6ex}

\end{titlepage}

%%%%%%%%%%%%%%%%%%%%%%%%%%%%%%%%%%%%%%%%

\section{Introduction}
\setcounter{equation}{0}

%%%%%%%%%%%%%%%%%%%%%%%%%%%%%%%%%%%%%%%%

Cosmic strings are macroscopic topological defects that may have been
produced at phase transitions in the early universe
\c{Kibble 76,Vilen 85}. They are predicted by gauge theories with
spontaneous breakdown of symmetry whenever the unbroken subgroup
contains a discrete symmetry as is the case of many grand unified
theories (GUT), although not in the simplest unified group $SU(5)$. The
GUT strings would have been produced when the universe was $10^{-34}
{\rm sec}$ old and had a temperature of $T \!\sim\! 10^{14-16} {\rm
GeV} $. The gravitational field of such strings may seed structure and,
in fact, a network of these strings is an alternative to inflation for
the generation of the universe structure \c{Vilen & Shell book 94,
Shellard 94,Vach & Vilen 91,Brand 85}.

There are two types of gravitational quantum effects associated to
cosmic strings (or to any other topological defect) which result from
the interaction of the string's gravitational field with any quantum
field ({\it i.e.}\ matter) present, namely, particle creation and
vacuum polarization. When a string forms, a sudden change in the
gravitational field takes place which may translate into copious
quantum pair production of particles in a way similar as
electron-positron pairs are created by external electromagnetic fields.
This effect has been considered by several authors in different
settings which go from various models of string formation
\c{Parker 87,Sahni 88,Mendell & His 89,Husain et al 90} to oscillating
string loops \c{Garriga et al 90}. The main conclusion is that even
though very energetic particles may be created the cosmological
significance of these is very small compared to the background
radiation at the time of formation. Typically, the ratio of the energy
density of particles created by the formation of the strings and the
energy density of the radiation is of order $N^{2}(G \mu )^{4}$, where
$N$ is the number of particle species, $G$ Newton's gravitational
constant, and $\mu$ is the energy per unit length of the string; for
GUT strings $\mu \!\sim\! 10^{22} {\rm g/cm}$, the square of the GUT
mass, and thus $G \mu \!\sim\! 10^{-6}$. For particles created by
oscillating loops such ratio is much higher, $N^{3/2} (G \mu)^{2}$,
but still cosmologically insignificant unless, of course, the number
of particles is absurdly large; at GUT time we expect $N \sim
10^{2}$.

Vacuum polarization effects due to quantum fluctuations of matter
fields have been much less studied. This is due, in part, to the fact
that one does not expect very significant changes in the classical
string network picture as a consequence of such effects,
but also in part because the computation of such effects is difficult
\c{His 87,Mazz & Lousto 91}. The vacuum expectation value of the stress
tensor of quantum fields around the string is generally different from
zero even for a static string, consequently it is the source of a
gravitational field which, in turn, modifies the classical
gravitational field of the string. This is the back-reaction effect of
quantum matter on the classical gravitational field. For example, the
gravitational field outside an unperturbed ({\it i.e.}\ not wiggly)
straight static string is described by the metric given by flat
spacetime with a deficit angle in the plane perpendicular to the string
\c{Vilen 81,Gott III 85,His 85,Garfinkle 85,Linet 85,Ford & Vilen 81}.
The quantum stress tensor for conformally coupled scalar fields has
been computed exactly by Linet \c{Linet 87} and by Helliwell and
Konkowski \c{Helliw & Konk 86} who found that the energy density goes
like $N \hbar \, G \mu \,r^{-4}$, where $r$ is the radial distance from
the string axis. Such energy density creates in the weak field
approximation a Newtonian potential outside the string of the order of
$\Phi \!\sim\! N \hbar \, G \mu \,r^{-2}$. In fact, Hiscock \c{His 87}
solved the semiclassical Einstein's equations to the linear order in
this case and found the back reaction in the gravitational field of
the cosmic string. The result is that the spacetime surrounding the
string is no longer flat with a deficit angle: the two-surface
perpendicular to the string is an hyperboloid (rather than a cone) and
the corrections to the flat metric are of the order just described. Two
consequences of this are clear, one is that a static string will exert
Newtonian forces on surrounding non relativistic particles, the other
is that when two cosmic strings approach they should feel increasingly
strong attractive forces. The relative significance of these effects
will be discussed later on.

In this paper we compute the back reaction on the gravitational field
of a cosmic string when it is formed. Now besides pure vacuum
polarization effects, as in the case of a static string, we have an
effect due to the creation of particles which also contributes to the
quantum stress tensor. The computation is made within the linear
approximation of the semiclassical correction to Einstein's equations,
{\it i.e.}\ we assume that the spacetime metric departs from the
Minkowski metric by linear terms $h_{\mu \nu}$ only. This approximation
is appropiate whenever the energy per unit length of the string
satisfies that $G \mu \ll 1$, which is certainly true for GUT strings.
Also the vacuum expectation value of the stress tensor for matter
fields is computed perturbatively to first order in $h_{\mu \nu}$ and
to the one loop order using the results by Horowitz \c{Horowitz 80} and
Jordan \c{Jordan 87} when the background metric is flat. More general
results for the stress tensor in conformally flat backgrounds are known
\c{Horowitz & Wald 80-82,Starob 81,Campos & Verd 94} but here we ignore
the effect due to the cosmological expansion. To this order the stress
tensor does not include the energy of the particles created, which is
second order in $h_{\mu \nu}$, however particle creation effects such
as those due to transition elements from vacuum to two particles states
are included in this linear order, see the discussion in ref.\
\c{Campos & Verd 94}.

To implement string formation we use a very simple macroscopic model
for the classical stress tensor of the string. The field dynamics of
string formation is a too complicated issue to be described at
microscopic level and one is restricted to consider rough macroscopic
models for this process. The models are either implemented by giving the
gravitational field of the string \c{Parker 87,Sahni 88,Mendell & His
89,Davies & Sahni 88} or by giving a prescribed form of the string
stress tensor \c{Husain et al 90,Garriga et al 90,Magueijo}.
The results for particle creation are rather independent of specific
models and thus we take the simplest model based on the stress tensor
given in ref.\ \c{Garriga et al 90}. In this model the string is always
there, first as a dust straight line source and it is the string
tension what grows in time, in this way the stress tensor is
automatically conserved.

The plan and a summary of the main results of the paper are the
following. In section 2 we describe the model of string formation and
derive its gravitational field in the linear approximation. We allow
for the possibility that the string be a straight wiggly string.
Straight wiggly strings are segments of long strings with small scale
structure whose effective stress tensor may be described by a straight
string with effective mass per unit length larger than the unperturbed
one  ({\it i.e.}\ with no small scale structure) and an effective
tension which is also less than the mass per unit length \c{Vilen 90}.
They appear as long strings in the numerical simulations of string
evolution \c{Bennet & Bouchet 90,Allen & Shell 90} and may be the
seeds of large scale structure \c{Vach & Vilen 91,Vollick 92,Copeland
et al 92}. As a consequence of linearity we can split the stress tensor
in two parts, one which is static and whose gravitational field is
easily solved and a time dependent part, whose gravitational field is
found by solving the corresponding initial value problem. For
simplicity we also use the sudden approximation, this introduces
divergences in the gravitational field along the future light cone
which must be regularized. To get rid of gauge effects which appear in
the metric tensor we also compute the Riemann tensor. The evolution at
large times of this tensor differs, of course, if the string is wiggly
or unperturbed. In the first case we have Newtonian like potentials
which help in building wakes \c{Vach & Vilen 91} whereas in the second
the curvature tends to zero corresponding to flat space with a deficit
angle.

In section 3 the vacuum expectation value of the stress tensor outside
the string for matter conformally coupled to the string's time
dependent gravitational field is derived. It is seen that after a
short transient period the stress tensor settles down to the values
of a static, wiggly or unperturbed, string. In both cases, the energy
density goes like $N \hbar \, G \!\mu \,r^{-4}$ as one expects from
dimensional arguments. Note that, for GUT strings, the string radius is
$r_{0}\!\sim\! 10^{-30} {\rm cm}$, corresponding to the Compton
wavelength of the GUT mass and this radius set bounds on the energy
density.

In section 4 the semiclassical Einstein's equations for our problem are
solved to find the back reaction of the quantum matter on the
gravitational field of the string. We can also split the stress tensor
into a static source, which includes the static part of the classical
source and of the quantum stress tensor, and a time dependent part. The
Riemann tensor is computed using some approximations in the stress
tensor and we discuss specially its behavior at large times. The
results are quite different if the string is wiggly or unperturbed. For
wiggly strings there is a Newtonian potential at the classical level
already and the quantum correction to such potential is too small to
quantitatively modify wake formation behind a moving string. For
unperturbed strings there is no Newtonian potential at the classical
level but a Newtonian potential appears as a quantum correction and,
thus, the string will exert a force on non relativistic matter.
However this effect would be only significant when $r$ is very small,
say of the order of $r_{0}$, {\it i.e.}\ only at a microscopic scale.
Perhaps back reaction might be significant to modify the dynamics of
string crossing, however, we should keep in mind that when two strings
approach at a distance of order $r_{0}$ the dynamics is dominated by
the microscopic field dynamics and the effective macroscopic picture
that we use breaks down.

%%%%%%%%%%%%%%%%%%%%%%%%%%%%%%%%%%%%

\section{Classical analysis}
\setcounter{equation}{0}

%%%%%%%%%%%%%%%%%%%%%%%%%%%%%%%%%%%%

In this section we compute the gravitational field created by the
string as it is formed. We use the following classical stress tensor
for the formation of a straight string which lies along the $z$-axis
\c{Garriga et al 90}
\be
T_{\!c \;\: \nu}^{\,\mu}=\mu \delta(x)\delta(y)\: {\rm diag}(1,0,0,
\tau\theta (t)), \label{class.s.t.}
\ee
where $\mu$ is the mass per unit length of the string, $\theta (t)$
the step function and $\tau$ a parameter ($0\!<\! \tau\! \leq\!1$) used
to modulate the tension, $\mu\tau$, of the string. If $\tau\!=\!1$ the
string is an unperturbed cosmic string whereas if $\tau\!<\!1$, this
tensor gives a macroscopic description of a straigth but wiggly cosmic
string. In this last case the effective equation of state for the
string is \c{Vilen 90,Carter 90} $\tau\mu^{2}\!=\! \mu_{ 0}^{2}$, where
$\mu_{ 0}$ is the unperturbed mass per unit length ({\it i.e.}\ for a
GUT string  $G \mu_{ 0} \!\sim\! 10^{-6}$). From numerical simulations
one has in the matter era the typical value \c{Vach & Vilen 91}
$\mu \!\sim\! 1.4 \mu_{ 0}$ ({\it i.e.}\ $\tau\!\sim\!0.5$). Note that
when the string forms each segment may be approximated by an
unperturbed cosmic string, a long segment becomes wiggly only after
evolution of the string network by string intersection and by chopping
off small loops. Thus back reaction is more important when an
unperturbed string forms, since it takes the smallest time, however we
keep $\tau$ arbitrary in what follows since vacuum polarization will
also exist once the string settles into a wiggly string and, in any
way, keeping $\tau$ arbitrary allows a simple identification ot the
time dependent effects.

Note also that in the classical stress tensor (\ref{class.s.t.}) we are
using two simultaneous approximations. The first is the thin line
approximation which assumes that the string has zero thickness, but, as
we have emphasized earlier, a physical string has a radius $r_{0}$
which gives a cut off radius inside which the approximation cannot be
trusted. The second is the step approximation which assumes that the
string is suddenly formed at time $t\!=\!0$, the use of $\theta (t)$
instead of a smooth function which grows from zero to one in a certain
time $T$ requires a cut off in the momentum of the quantum modes of
the order of $1/T$. The time of formation $T$ is bounded by the age of
the universe when the string forms (for GUT strings $T\!<\!10^{-34}
{\rm sec}$).

In the weak field approximation the metric in Cartesian coordinates
can be written as
\be
g_{\mu\nu}(x)=\eta_{\mu\nu}+h_{\mu\nu}(x),   \label{metric}
\ee
where $\eta_{\mu\nu}\!=\!{\rm diag}(1,-1,-1,-1){\rm,}\;
\mid\!h_{\mu\nu}\!\mid\ll\!1$ and we have the gauge freedom, due to
infinitessimal coordinate changes $x'^{\mu}\!=\!x^{\mu}+
\xi^{\mu}(x)$, that
\be
h'_{\mu\nu}=h_{\mu\nu}+\xi_{\mu,\nu}+\xi_{\nu,\mu}
\ee
for arbitrary $\xi^{\mu}(x)$.
Einstein's equations for the metric perturbation $h_{\mu\nu}$ can be
written in the harmonic gauge $(h^{\mu\nu}\!\!-\!\frac{1}{2}\,
\eta^{\mu\nu}h^{\alpha}_{\;\:\alpha}),_{\,\nu}=0$ as
\be
\Box h_{\mu\nu}=-16\pi\, G S_{\mu\nu},   \label{Einstein eqs}
\ee
where $S_{\mu\nu}\equiv T_{\mu\nu}-\!\frac{1}{2}\,\eta_{\mu\nu}
T^{\alpha}_{\;\:\alpha}$ and we still have the freedom within the
harmonic gauge of choosing functions $\xi^{\mu}(x)$ such that
$\Box \xi^{\mu}(x)\!=\!0$. Note that a different analysis of this model
which goes beyond the linear approximation has beeen given in ref.\
\c{Magueijo}.

Since we start with a static prescribed metric before the string forms
at $t\= 0$ we can find the solution of (\ref{Einstein eqs}) for $t>0$
as a Cauchy problem. As it is well known, such a solution is given by
Kirchhoff's formula,
\begin{eqnarray}
h_{\mu\nu}(t,\vec{x}\,)&=&\int_{_{\displaystyle I\hspace{-0.7 ex}
R^{3} }} \!\!\!\!
d^{3}\vec{x\,}' \:\left[\partial_{t'}h_{\mu\nu}(0,\vec{x\,}') \,
D(t,\vec{x}-\vec{x\,}')
+h_{\mu\nu}(0,\vec{x\,}') \,\partial_{t} D(t,\vec{x}-\vec{x\,}')
\right] \nonumber \\
&&-16 \pi G\int_{0}^{t} dt' \int_{_{\displaystyle I\hspace{-0.7 ex}
R^{3}}}
 \!\!\!\! d^{3}\vec{x\,}' \: D_{R}(t-t',\vec{x}-\vec{x\,}') \,
S_{\mu\nu}(t',\vec{x\,}'),
\label{Kirchhoff}
\end{eqnarray}
where $D_{R}$ is the retarded Green function for a massless scalar
field, {\it i.e.}
$D_{R}(x\!-\!x')\!=\!\frac{1}{2\pi}\, \delta[(x\!-\!x')^{2}] \,
\theta(t\!-\!t')$ or
$\frac{1}{4\pi}\, \delta(t\!-\!t'\!-\!\mid
\!\vec{x}\!-\!\vec{x\,}'\!\mid) / \!\!\mid \!\vec{x}\!-\!\vec{x\,}'\!
\mid$, and
$D\!=\!D_{R}\!-\!D_{A}$ is the Schwinger Green function ($D_{A}$ is
the advanced Green
function), which may be written as $D(x\!-\!x')\!=\! \frac{1}{2\pi}\,
\delta[(x\!-\!x')^{2}] \,[\theta(t\!-\!t')\!-\!\theta(t'\!-\!t)]$.
The first integral in (\ref{Kirchhoff}) is over the hypersurface
$t'\!=\!0$ and it
is the solution of the homogenous equation which satisfies the boundary
conditions given by the metric, $h_{\mu\nu}$, and its first time
derivative, $\partial_{t'}h_{\mu\nu}$, at $t'\=0$. The second
integral is the solution of the inhomogeneous equation which vanishes
at $t\=0$ ({\it i.e.}
the boundary conditions are implemented with the solution of the
homogeneous equation) and has support,
see $D_{R}$, on the truncated past light cone starting at $(t,\vec{x})$
and ending at the hypersurface $t'\=0$.

The solution of (\ref{Einstein eqs}) for the stress tensor
(\ref{class.s.t.}) is somewhat simplified if we write
(\ref{class.s.t.}) as the sum of two stress tensors, both conserved,
one of them, $T^{{\scriptscriptstyle (1)}}_{\mu\nu}$, static and the
other, $T^{{\scriptscriptstyle (2)}}_{\mu\nu}$, time dependent,
{\it i.e.}
$T_{\mu\nu}\=T^{{\scriptscriptstyle (1)}}_{\mu\nu}\+
T^{{\scriptscriptstyle (2)}}_{\mu\nu}$,
with
\be
T_{\;\;\;\;\;\;\; \nu}^{{\scriptscriptstyle  (1)}\,\mu}=\mu
\delta(x)\delta(y)\: {\rm diag}(1,0,0,0),\;\;\;\;
T_{\;\;\;\;\;\;\; \nu}^{{\scriptscriptstyle  (2)}\,\mu}=\mu \tau
\delta(x) \delta(y) \,\t (t)\: {\rm diag}(0,0,0,1).
\ee
This leads to write
$h_{\mu\nu}\=h^{{\scriptscriptstyle (1)}}_{\mu\nu}\+
h^{{\scriptscriptstyle (2)}}_{\mu\nu}$,
where $h^{{\scriptscriptstyle (1)}}_{\mu\nu}$ is the static metric
corresponding to the
source $T^{{\scriptscriptstyle (1)}}_{\mu\nu}$ and
$h^{{\scriptscriptstyle (2)}}_{\mu\nu}$
is the time dependent part corresponding to the source
$T^{{\scriptscriptstyle (2)}}_{\mu\nu}$. The field equation
(\ref{Einstein eqs}) for the static part
$h^{{\scriptscriptstyle (1)}}_{\mu\nu}$ is now
simply
\be
\nabla^{2} h^{{\scriptscriptstyle (1)}}_{\mu\nu}=8\pi G \mu
\delta(x)\delta(y)\: {\rm diag}(1,1,1,1)
\ee
which, using the cylindrical symmetry of the problem, has the solution
\be
h^{{\scriptscriptstyle (1)}}_{tt}=h^{{\scriptscriptstyle (1)}}_{xx}=
h^{{\scriptscriptstyle (1)}}_{yy}=h^{{\scriptscriptstyle (1)}}_{zz}=
4 G \mu \,\ln \!\left(\frac{r}{R} \right) \equiv \alpha (r),
\label{alpha}
\ee
where $r^{2}\=x^{2}\+y^{2}$ and $R$ is an arbitrary constant with
dimensions of length.

The field equation for the time dependent metric perturbation
$h^{{\scriptscriptstyle (2)}}_{\mu\nu}$ is now given by
\be
\Box h^{{\scriptscriptstyle (2)}}_{\mu\nu}=-8\pi G \mu \tau
\delta(x)\delta(y)\,\theta(t) \: {\rm diag}(-1,1,1,-1)
\ee
whose solution as a Cauchy problem is given by (\ref{Kirchhoff}), but
we note that since the boundary conditions for these components are
obviously
\mbox{$h^{{\scriptscriptstyle (2)}}_{\mu\nu}(0,\vec{x})\=0$},
\mbox{$\partial_{t}h^{{\scriptscriptstyle (2)}}_{\mu\nu}(0,\vec{x})
\=0$} we are left with the second integral of (\ref{Kirchhoff}) only.
Using the second representation for the retarded Green function given
above we have after a simple integration that the (unique)
solution for $h^{{\scriptscriptstyle (2)}}_{\mu\nu}$ is
\be
h^{{\scriptscriptstyle (2)}}_{tt}=h^{{\scriptscriptstyle (2)}}_{zz}=
-h^{{\scriptscriptstyle (2)}}_{xx}=-h^{{\scriptscriptstyle (2)}}_{yy}=
4 G \mu\tau \,{\rm arccosh}\! \left( \frac{t}{r} \right)\,\theta(t-r)
\equiv \beta (t,r).
\label{beta}
\ee
It is easy to check that the final metric perturbation
$h_{\mu\nu}\=h^{{\scriptscriptstyle (1)}}_{\mu\nu}\+
h^{{\scriptscriptstyle (2)}}_{\mu\nu}$
satisfies the harmonic gauge condition. We can write the gravitational
field of the string in cylindrical coordinates $(t,r,\theta,z)$ as
\be
ds^{2}=(1+\alpha+\beta)\,dt^{2}-(1-\alpha-\beta)\,dz^{2}-
(1-\alpha+\beta)\,(dr^{2}+r^{2}
d\theta^{2}),     \label{class.metric}
\ee
where $\alpha(r)$ and $\beta(t,r)$ are given by (\ref{alpha}) and
(\ref{beta}) respectively.

A few comments on the metric (\ref{class.metric}) are now in order.
First, we see that the metric is continuous but its first derivatives
are discontinuous along the light cone $t\=r$. This is a consequence
of the use of the step approximation: had we used a smooth time
dependent function instead of $\theta(t)$ the derivatives would be
smooth too. This means that we can only trust our results outside the
spacetime region bounded by $t\= r\-T/2$ and $t\= r\+T/2$. To this we
should add that due to the thin line approximation the results are only
reliable for $r\!>\!r_{0}$, where $r_{0}$ is the string radius.

Second, we also note that the metric perturbations diverge at
$r\!\rightarrow\!\infty$ ($t$ fixed) and at  $t\!\rightarrow\!\infty$
($r$ fixed) but these divergences, as we will see from the Riemann
tensor, are only gauge effects, they are a consequence of the use of
the harmonic gauge.

Third, one expects that when $t\gg r$ ($r$ fixed) the metric becomes
that of a static cosmic string, unperturbed if $\tau\=1$ or wiggly
otherwise. But this again cannot be seen directly in the harmonic
gauge. The coordinate change which puts the above metric in a
suitable form to see this is rather messy and, since it gives no
further light on this issue, we shall not write it here.

An important physical observable is the deficit angle of the
two-surfaces $t\={\rm const.}$,
$z\={\rm const.}$ for the metric (\ref{class.metric}). Following ref.\
\c{Ford & Vilen 81}, given a closed piecewise smooth curve on a
two-surface which encloses a regular and simply connected region $S$,
one defines the deficit angle associated with this curve as the angle
$\triangle \t$ rotated by a vector after being parallel transported
once around the curve. An application of the Gauss-Bonnet theorem shows
that this angle is given by the surface integral of the Gaussian
curvature $K$ over $S$: $\triangle \t = \int_{S }  K \, dS$. If the
two-surface has circular symmetry the deficit angle is defined as the
deficit angle associated with circles of radius $r$. It is easy to see
that in a space-time described by a cylindrical symmetric metric the
deficit angle of the two-surfaces $t\={\rm const.}$, $z\={\rm const.}$
reduces to
\be
\triangle \t (t,r)= 2 \pi \left[1-\frac{1}{\sqrt{-g_{rr}}} \,
\frac{\!\!\!\p}{\p r} \sqrt{-g_{\t\t}} \right],  \label{def angle}
\ee
which in the linear approximation becomes
$\triangle \t (t,r)=  \pi \left( h^{r}_{r}-h^{\t}_{\t}-
r\,h^{\t}_{\t,r} \right)$. Note that this definition differs from that
used in ref.\ \c{His 87}. For the metric (\ref{class.metric}) we obtain
\be
\triangle \t (t,r)=  4\pi G\mu  \left[1+\tau \,
\frac{t}{ \sqrt{t^{2}-r^{2}}}\,\theta(t\-r) \right].
\ee
It is clear that when $t\!\gg\!r$, $r$ fixed,
$\triangle \t\!\rightarrow\! 4\pi G\mu \, (1+\tau)$, which is the
deficit angle for a static string.

The properties of the metric (\ref{class.metric}) are better
deduced from the Riemann tensor which is gauge
independent. Note also that some
Riemann components give the tidal forces on surrounding test particles.
In the linear approximation we have
\be
R_{\mu\nu\alpha\beta} =\frac{1}{2}\,
(\d^{\r}_{\mu}\d^{\s}_{\b}\p_{\nu}\p_{\a}+\d^{\r}_{\nu}\d^{\s}_{\a}
\p_{\mu}\p_{\b}-
\d^{\r}_{\nu}\d^{\s}_{\b}\p_{\mu}\p_{\a}-\d^{\r}_{\mu}\d^{\s}_{\a}
\p_{\nu}\p_{\b})
\, h_{\r\s}.   \label{Riemann}
\ee
Following the separation of the metric perturbation into a static part
and a time dependent part,
$h_{\mu\nu}\=h^{{\scriptscriptstyle (1)}}_{\mu\nu}\+
h^{{\scriptscriptstyle (2)}}_{\mu\nu}$,
the Riemann components can also be separated as
$R^{{\scriptscriptstyle (1)}}_{\mu\nu\alpha\beta}$ and
$R^{{\scriptscriptstyle (2)}}_{\mu\nu\alpha\beta}$, the last is easily
identified because it is proportional to $\tau$. For $r\!\neq\!0$ they
are
\begin{eqnarray}
R_{\,tztz}&=&-R_{\,xyxy}=\frac{2G\mu \tau }{r^{2}}\;{\cal P}\!f
\left[\frac{b}{(b^{2}\-1)^{3/2}}\:
\t(b\-1) \right], \nonumber\\
R_{\,xzxz}&=&\frac{2G\mu}{ r^{2}}\;\left\{\cos 2\t-\tau \,{\cal P}\!f
\left[\frac{b}{\sqrt{b^{2}\-1}}\:\left(\cos 2\t-
\frac{\cos^{2}\t}{(b^{2}\-1)}\right) \t(b\-1)\right]\right\},
\nonumber\\
R_{\,xzyz}&=&R_{\,txty}=\frac{2G\mu}{r^{2}}\;\sin 2\t \;\left\{1-
\frac{\tau}{2}\,
{\cal P}\!f \left[\frac{b \,(2 b^{2}\-3)}{(b^{2}\-1)^{3/2}}\:\t(b\-1)
\right]\right\}, \nonumber\\
R_{\,txtx}&=&\frac{2G\mu}{r^{2}}\;\left\{\cos 2\t-\tau \,{\cal P}\!f
\left[\frac{b}{\sqrt{b^{2}\-1}}\:\left(\cos 2\t+
\frac{\sin^{2}\t}{(b^{2}\-1)}\right) \t(b\-1)\right]\right\},
\nonumber\\
R_{\,tzxz}&=&R_{\,tyyx}=-\frac{2G\mu \tau}{r^{2}}\;\cos \t \;
{\cal P}\!f \left[\frac{1}{(b^{2}\-1)^{3/2}}\:
\t(b\-1)\right],
\label{class.Riemann}
\end{eqnarray}
where we have introduced the new variable $b\!\equiv\!t/r$, instead of
$t$, and where ${\cal P}\!f$ denotes the Hadamard finite part, which
gives well defined expressions, in the sense of distributions, on the
light cone $b\=1$ (see Appendix A). $R_{\,yzyz}$, $R_{\,tyty}$,
$R_{\,tzyz}$ and $R_{\,txxy}$ can be obtained from $R_{\,xzxz}$,
$R_{\,txtx}$, $R_{\,tzxz}$ and $R_{\,tyyx}$ respectively by
interchanging $\cos \t$ and $\sin \t$.

It is now clear that when $\tau\=1$ (unperturbed string) the Riemann
tensor vanishes when $t\!\rightarrow\!\infty$ ($r$ fixed), {\it i.e.}\
$\lim_{t \rightarrow \infty} R_{\mu\nu\a\b}\=0$, and the spacetime
becomes flat. But if $\tau\!\neq\!1$ there are tidal forces among test
particles wich correspond to a Newtonian like potential
$h_{tt}\!\propto\!\ln r$. Note that a Riemann component such as
$R_{tztz}$, which gives relative accelerations among particles along
the direction of the string, is a transient term, {\it i.e.}\
$R_{tztz}\=0$ when $t\!<\!r$ (in regions not yet affected by the
formation of the string) and $R_{tztz} \simeq 2G\mu\tau \, r^{-2} \,
b^{-2} \left(1+\frac{3}{2}\, b^{-2}+O \left(b^{-4}\right)
\right)$ when $t\!>\!r$, which approaches zero very quickly.

Our next task is to obtain the quantum correction to this curvature
tensor due to the quantum fluctuations of matter fields.

%%%%%%%%%%%%%%%%%%%%%%%%%%%%%%%%%%%%%%%%%%%%%%%%%%

\section{The stress tensor for matter fields}
\setcounter{equation}{0}

%%%%%%%%%%%%%%%%%%%%%%%%%%%%%%%%%%%%%%%%%%%%%%%%%%

Quantum fluctuations of matter fields interacting with the
gravitational field of a cosmic string give a non null vacuum
expectation value for the stress tensor of these matter fields, even if
they are conformally coupled. For a free massless conformally coupled
scalar field in a flat spacetime background with arbitrary linear
gravitational perturbations (\ref{metric}) this stress tensor has been
computed to one loop order by several authors \c{Horowitz 80,Jordan 87,
Horowitz & Wald 80-82,Starob 81,Campos & Verd 94} and it is given,
to first order in $h_{\mu\nu}(x)$, by
\be
\langle T^{\mu\nu}(x) \rangle=-\frac{\a \hbar}{6}\,B^{\mu\nu}(x)+3 \a
\hbar \int d^{4}y\:
H(x\-y,\bar{\mu})\, A^{\mu\nu}(y),  \label{quant.s.t.}
\ee
where $\a\!\equiv\!(2880 \pi^{2})^{-1}$,
\begin{eqnarray}
B^{\mu\nu}(x)&=&2 \eta^{\mu\nu} \Box G^{\;\: \a}_{\!\a}-2 G^{\;\: \a,
\mu\nu}_{\!\a},\nonumber\\
A^{\mu\nu}(x)&=&-2 \Box G^{\mu\nu}-\frac{2}{3} G^{\;\: \a,\mu\nu}_{\!
\a}+
\frac{2}{3} \eta^{\mu\nu} \Box G^{\;\: \a}_{\!\a},   \label{A & B}
\end{eqnarray}
$G^{\mu\nu}$ is the Einstein tensor for the metric to first order in
$h_{\mu\nu}$, and
$H(x\-y,\bar{\mu})$ is a propagator defined by
\be
H(x\-y,\bar{\mu})\equiv -\frac{1}{2} \int \frac{d^{4}p}{(2\pi)^{4}}\:
e^{-i p (x-y)}\,
\left[\ln\left(\frac{-(p^{2}+i\epsilon)}{\bar{\mu}^{2}}\right)+2 \pi i
\, \t(p^{2}) \t(-p^{0})\right], \label{H}
\ee
where $\bar{\mu}$ is an arbitrary renormalization scale. Notice that
the second
term in (\ref{quant.s.t.}) is traceless and that it is the term
proportional to
$B^{\mu\nu}(x)$ which gives the trace anomaly in this case,
\be
\langle T^{\mu}_{\;\:\mu}(x) \rangle=-\frac{\a \hbar}{6}\,
B^{\mu}_{\;\:\mu}(x)=
\a \hbar \,\Box R,
\ee
where $R$ is the scalar curvature to first order in $h_{\mu\nu}$. Using
$\nabla_{\!\mu} G^{\mu\nu}\=0$, it is easily seen from (\ref{A & B})
that the two
terms in (\ref{quant.s.t.}) are independently conserved.

We can now proceed to the computation of
$\langle T^{\mu\nu}(x) \rangle$ to first order in $h_{\mu\nu}$. In
section 3.1 we will calculate it outside the string, that is, for
$r\!\neq\! 0$. In section 3.2 we will see how a generalization to
include $r\=0$ can be made. In section 3.3 we give the approximation
to the stress tensor which will be finally used to compute the
back reaction on the string metric.

%%%%%%%%%%%%%%%%%%%%%%%%%%%%%%%%%%%%%%%%%%%%%%%%%%%%%%%%%%%%%

\subsection{The quantum stress tensor outside the string}

%%%%%%%%%%%%%%%%%%%%%%%%%%%%%%%%%%%%%%%%%%%%%%%%%%%%%%%%%%%%%

{}From (\ref{quant.s.t.}) and (\ref{A & B}) we see that all the
dependence on the metric perturbations is in the Einstein tensor
$G_{\mu\nu}$. Now since the source of such gravitational perturbations
is the classical string source $T_{\!c}^{\,\mu\nu}$ given in
(\ref{class.s.t.}) we can use Einstein's equations $G^{\mu\nu}\=-8\pi
G \,T_{\!c}^{\,\mu\nu}$ in (\ref{A & B}). Thus it is worth to remark
that we do not need to know the explicit gravitational field created by
the forming string, it suffices to know the explicit form of the
classical stress tensor which produces that field. After the above
substitution, since $T_{\!c}^{\,\mu\nu}$ is proportional to
$\d(x)\d(y)$, both $B^{\mu\nu}(x)$ and $A^{\mu\nu}(x)$ have support on t
he string core. If we are interested in computing $\langle
T^{\mu\nu}(x) \rangle$ outside the string, {\it i.e.}\
for $r\!\neq\! 0$, it is clear from (\ref{quant.s.t.}) that the only
contribution comes from the second term in (\ref{quant.s.t.}) and that
terms in $H(x\-y,\bar{\mu})$ with support on $x\=y$ will not contribute.

Let us give a more suitable representation of $H(x\-y,\bar{\mu})$ for
$x\!\neq\! y$. For this it is useful to define the four-vector
\be
F_{\a}(x\-y)\equiv -\frac{1}{2} \int \frac{d^{4}p}{(2\pi)^{4}}\:
e^{-i p (x-y)}\,
\frac{\!\!\!\!\!\p}{\p p^{\a}} \left[ \ln\left(
\frac{-(p^{2}+i\epsilon)}{\bar{\mu}^{2}}\right)
+2 \pi i \,\t(p^{2}) \t(-p^{0})\right].   \label{F}
\ee
This vector can, on the one hand, be simply computed as
\begin{eqnarray}
F_{\a}(x\-y)&=& -i\,\frac{\!\!\!\!\!\p}{\p x^{\a}} \int
\frac{d^{4}p}{(2\pi)^{4}}\: e^{-i p (x-y)}\,
\left[\frac{1}{p^{2}+i \epsilon}+2 \pi i \, \d(p^{2}) \t(-p^{0})
\right] \nonumber\\
&=&i\, \frac{\!\!\!\!\!\p}{\p x^{\a}} D_{R}(x-y)=\frac{i}{\pi}\,
(x-y)_{\a} \,\d '\!\left[ (x-y)^{2}\right]
\t(x^{0}-y^{0}),  \label{F2}
\end{eqnarray}
where $\d '$ means derivative with respect to the argument of $\d$
and we have used a well known integral representation of the retarded
Green function $D_{R}$ and the form of $D_{R}$ given in section 2.

On the other hand, integrating (\ref{F}) by parts, we can also write
$F_{\a}(x\-y)$ as,
\be
F_{\a}(x\-y)= i \,(x-y)_{\a}\, H(x\-y,\bar{\mu}).  \label{F3}
\ee
Now comparing (\ref{F2}) and (\ref{F3}) we
get the following useful representation for $H$, when $x\!\neq\! y$,
\be
H(x\-y,\bar{\mu})= \frac{1}{\pi} \, \d'\! \left[(x-y)^{2} \right]
\t(x^{0}-y^{0}).    \label{propagator}
\ee

Thus using (\ref{propagator}) we have outside the string
\be
\langle T^{\mu\nu}(x) \rangle= \frac{3 \a \hbar}{\pi} \int d^{4}y
\:\d'\! \left[(x-y)^{2} \right]
\t(x^{0}-y^{0}) A^{\mu\nu}(y),  \label{quant.s.t.2}
\ee
with
\be
A^{\mu\nu}(x)=16 \pi G \,\eta^{\a\b\mu\nu}_{\:\:\:\:\:\:\:\:\:\:\,
\r\s}\,\, \p_{\a}\p_{\b}
T_{\!c}^{\,\r\s},
\ee
where we have defined
\be
\eta^{\a\b\mu\nu}_{\;\;\;\;\;\;\;\;\, \r\s} \equiv \eta^{\a\b}
\d^{\mu}_{\r} \d^{\nu}_{\s}
+\frac{1}{3}\, \eta_{\r\s}\, (\eta^{\mu\a} \eta^{\nu\b}-\eta^{\mu\nu}
\eta^{\a\b}).
\ee
If we now write, see ref.\ \c{Jordan 87}, ${\displaystyle \d'\!
\left[(x\-y)^{2} \right]\=\lim_{\lambda
\rightarrow 0^{-}} \d'\! \left[(x\-y)^{2}+\lambda \right]\=
\lim_{\lambda \rightarrow 0^{-}}
\frac{\!\!d}{d\lambda} \d\! \left[(x\-y)^{2}+\lambda \right]}$ in
(\ref{quant.s.t.2}),
change variables and use (\ref{class.s.t.}), we have explicitly
\begin{eqnarray}
\langle T^{\mu\nu}(x) \rangle &=& 48 \a G \mu \hbar \,
\eta^{\a\b\mu\nu}_{\;\;\;\;\;\;\;\;\, \r\s}\,
\frac{\!\!\!\!\!\p}{\p x^{\a}}
\frac{\!\!\!\!\!\p}{\p x^{\b}}
\,\lim_{\lambda \rightarrow 0^{-}}
\frac{\!\!d}{d\lambda}\, \int d^{4}x'\,\d(x'^{2}+\lambda) \t (t')
\nonumber\\
&& \hspace*{12.3ex} \times \d(x-x') \d(y-y') \left( \d^{\r}_{t}
\d^{\s}_{t}-\tau \t(t-t')\, \d^{\r}_{z} \d^{\s}_{z} \right)
\label{quant.s.t.3}
\end{eqnarray}
and the computation is now straightforward. It is clear from
(\ref{quant.s.t.3}) that we have
two types of terms in the stress tensor, $\langle T_{\mu\nu}\rangle\=
\tilde{T}^{{\scriptscriptstyle (1)}}_{\!A \,\mu\nu}+
\tilde{T}^{{\scriptscriptstyle (2)}}_{\!A \,\mu\nu}$, the static terms,
$\tilde{T}^{{\scriptscriptstyle (1)}}_{\!A \,\mu\nu}$, which depend on
the static part of
$T_{\!c}^{\,\mu\nu}$ and the time dependent terms,
$\tilde{T}^{{\scriptscriptstyle (2)}}_{\!A \,\mu\nu}$, which depend on
the time dependent part, that is, the terms which are proportional to
$\tau \t(t\-t')$. The subindex ${\scriptstyle A}$ stands for the part
of the stress tensor related to the tensor $A^{\mu\nu}$ in
(\ref{quant.s.t.}). The static terms are easily
computed, the $\int dx'\,dy'$ integration is trivial and we use
$\d(x'^{2}+\lambda)$ to perform
the $\int dt'$ integration, following ref.\ \c{Campos & Verd 94}.
These terms are all proportional to $1/r^{4}$. The time dependent
terms, although sligthly more complicated, can be computed without
difficulty. The final result for $r\!\neq\!0$ in the Cartesian
coordinate basis is:
\begin{eqnarray}
&&\!\!\!\!\!\!\!\!\!\langle T_{t}^{t}\rangle = -\frac{2 \s}{r^{4}}
\left[4- \tau \,f_{18}(b)
                             \right], \;\;\;\;\;\,
  \langle T_{z}^{z}\rangle = \frac{4 \s}{r^{4}} \left[ 1- \tau\,
f_{12}(b)
                             \right],   \nonumber\\
&&\!\!\!\!\!\!\!\!\!\langle T_{t}^{a}\rangle = \frac{3 \s \, x^{a}}
{r^{5}}\, \tau\, g(b),   \;\;\;\;\;\;\;\;\;\;\;\;\;\;\;
  \langle T_{b}^{a}\rangle = \frac{\s}{r^{4}}\left\{ 3 \d^{ab}
\left[ 2+ \tau \,f_{12}(b)\,
                             \right] - 4 \frac{x^{a} x^{b}}{r^2}
\left[ 2+\tau\, f_{15}(b)
                             \right] \right\} ,\nonumber\\
         \label{quant.s.t.4}
\end{eqnarray}
where $\s \= 8\a G \mu \hbar$, the index $a\=1,2$ refers in $x^{a}$ to
$x^{1}\=x$, $x^{2}\=y$ (the string transversal coordinates), and
\be
g(b)\equiv {\cal P}\!f \left[\frac{1}{(b^{2}-1)^{5/2}} \,\t(b\-1)
\right] ,
\;\; f_{C}(b) \equiv  {\cal P}\!f \left[\frac{b}{(b^{2}-1)^{5/2}}\,
(2b^{4}-5b^{2}+C/4)\,\t(b\-1)\right].
\label{g f}
\ee
As above ${\cal P}\!f$ denotes the Hadamard finite part, which gives
well defined distributions on the light cone $b\=1$ (see Appendix A).
It is easy to identify directly from (\ref{quant.s.t.4}) the static
and time dependent parts of the stress tensor, {\it i.e.}\
$\tilde{T}^{{\scriptscriptstyle (1)}}_{\!A \,\mu\nu}$ and
$\tilde{T}^{{\scriptscriptstyle (2)}}_{\!A \,\mu\nu}$ respectively.
Note that
$\tilde{T}^{{\scriptscriptstyle (1)}}_{\!A \,\mu\nu}$ corresponds to
the vacuum expectation value of the stress tensor due to a dust rod
along the $z$-axis with mass per unit length $\mu$. It can be checked,
computing the derivatives of the distributions (\ref{g f})
using the methods of Appendix A, that the $\langle T_{\mu\nu}\rangle$
just derived is conserved.

The most salient feature of $\langle T_{\mu\nu}\rangle$ is that it
quickly settles down to the final static values. Note that when $t$
grows keeping $r$ fixed, $b\!\equiv\!t/r$ grows as $t$ and we can
expand $g(b)$ and $f_{C}(b)$ in terms of $b^{-1}$: $g(b)\!\simeq\!
b^{-5} \left(1\+\frac{5}{2} \, b^{-2} \+O \left(b^{-4} \right) \right)$,
$\, f_{C}(b)\!\simeq\! 2 \left(1\+O \left(b^{-4} \right) \right)$.
This means that $f_{C}(b)$ differs from the static value, $2$, by
terms of order $b^{-4}$ and that $g(b)$ goes like
$b^{-5}$. Thus $\langle T^{\mu\nu}(x) \rangle$ is effectively time
dependent only when $t\!\sim\!r$, it reaches the static values very
quickly.

The final static values of $\langle T_{\mu\nu}\rangle$, {\it i.e.}\
when $t \rightarrow \infty$, are read off from (\ref{quant.s.t.4}). In
the polar coordinate basis $(\p_{t},\p_{r},\p_{\t},\p_{z})$ they are
\be
\lim_{t\rightarrow \infty} \langle T^{\mu}_{\;\:\nu}\rangle= -
\frac{16 \a G\mu \hbar}{r^{4}}
\: {\rm diag}(4-2\tau,1+\tau,-3-3\tau,-2+4\tau).
\ee
In particular when $\tau\=1$, {\it i.e.}\ the string is not wiggly, we
get the well known results of refs. \c{Linet 87,Helliw & Konk 86}.

%%%%%%%%%%%%%%%%%%%%%%%%%%%%%%%%%%%%%%%%%%%%%%%%%%%%%%%%%%

\subsection{The quantum stress tensor including $r\=0$}

%%%%%%%%%%%%%%%%%%%%%%%%%%%%%%%%%%%%%%%%%%%%%%%%%%%%%%%%%%

Since the quantum stress tensor is a source in the semiclassical
Einstein's equations, we need to know $\langle T_{\mu\nu}\rangle$ in
all the space-time. One way to do this is to try to generalize the
previous calculation to include $r\=0$. Such a generalization should
have the form
\be
\langle T^{\mu\nu}(x) \rangle=\tilde{T}_{\!B}^{\,\mu\nu}(x)+
\tilde{T}_{\!A }^{{\scriptscriptstyle (1)}{\mu\nu}}(x)+
\tilde{T}_{\!A }^{{\scriptscriptstyle (2)}{\mu\nu}}(x),
\label{quant.s.t.5}
\ee
where $\tilde{T}_{\!B}^{\,\mu\nu}\!\equiv\! -(\a \hbar/6)\,B^{\mu\nu}$
and $\tilde{T}_{\!A }^{{\scriptscriptstyle (1)}{\mu\nu}}$ and
$\tilde{T}_{\!A }^{{\scriptscriptstyle (2)}{\mu\nu}}$ are the traceless
tensors corresponding to the static and time dependent parts
respectively in the second term of (\ref{quant.s.t.}).
These should be some well defined distributions which reduce to the
expressions (\ref{quant.s.t.4}) for $r\!\neq\!0$. Using Einstein's
equations $G^{\mu\nu}\=-8\pi G \,T_{\!c}^{\,\mu\nu}$ in
(\ref{A & B}), $\tilde{T}_{\!B}^{\,\mu\nu}$ can be expressed in terms
of the classical stress tensor so its computation is straightforward.
One finds the following non-null components in the Cartesian
coordinate basis:
\begin{eqnarray}
&&\!\!\!\!\!\!\!\!\!\!\tilde{T}_{\!B \,t}^{\;\;\,t}= -\frac{\pi}{3}\,
\s \left( 1+\tau\,\t(t) \right)
                             \nabla^{2} \left(\d(x) \d(y) \right),
                             \;\;\;\;\;\;\;\;\;\;
\tilde{T}_{\!B \,t}^{\;\;\,a}=
                             \frac{\pi}{3}\, \s \tau \, \d(t)\, \p_{a}
\left(\d(x)\d(y)\right),  \nonumber\\
&&\!\!\!\!\!\!\!\!\!\!\tilde{T}_{\!B \,z}^{\;\;\,z}= -\frac{\pi}{3}\,
\s  \left[ \left( 1+\tau\,\t(t) \right)
                             \nabla^{2} \left(\d(x) \d(y) \right)-
\tau \, \d'(t) \,
                             \d(x) \d(y) \right],  \nonumber\\
&&\!\!\!\!\!\!\!\!\!\!\tilde{T}_{\!B \,b}^{\;\;\,a}= -\frac{\pi}{3}\,
\s  \Bigl\{  \d_{ab} \left[ \left( 1+\tau\,\t(t) \right)
                             \nabla^{2} \left(\d(x) \d(y) \right)-\tau
\, \d'(t) \,
                             \d(x) \d(y) \right]  \nonumber\\
&& \hspace*{10ex}- \left( 1+\tau\,\t(t) \right)
                             \p_{a}\p_{b}\left(\d(x)\d(y)\right)
                             \Bigr\}.
                             \label{quant.s.t.B}
\end{eqnarray}
It is easy to see that the tensor $\tilde{T}_{\!B}^{\,\mu\nu}$ just
found is conserved. From the expressions (\ref{quant.s.t.B}) it is
clear that we can write $\tilde{T}_{\!B}^{\,\mu\nu}\=
\tilde{T}_{\!B}^{\,{\scriptscriptstyle (1)}\mu\nu}+
\tilde{T}_{\!B}^{\,{\scriptscriptstyle (2)}\mu\nu}$ where, as always,
$\tilde{T}_{\!B}^{\,{\scriptscriptstyle (1)}\mu\nu}$ refers to the
static terms and $\tilde{T}_{\!B}^{\,{\scriptscriptstyle (2)}\mu\nu}$
to the time dependent terms.

To generalize the calculation of the second term in (\ref{quant.s.t.})
to include $r\=0$, we need to extend the representation
(\ref{propagator}) of the propagator $H(x\-y,\bar{\mu})$ to all values
of $(x\-y)^{\a}$. Such an extension, which is derived in Appendix A, is
\be
H(x-y,\bar{\mu})= \lim_{\epsilon \rightarrow 0^{+}} \left\{
\frac{1}{\pi}
\, \d'\! \left[(x-y)^{2} \right] \t(x^{0}-y^{0}) \,
\t \left(\mid \!\vec{x}\-\vec{y} \!\mid \! -\epsilon \right)+
\left[ \ln \bar{\mu} \epsilon+\gamma \-1 \right]  \d^{4}(x-y) \right\},
\label{propagator2}
\ee
where $\gamma$ is Euler's constant. For the static terms
$\tilde{T}_{\!A }^{{\scriptscriptstyle (1)}{\mu\nu}}$ we find
\begin{eqnarray}
&&\!\!\!\!\!\!\!\!\!\!\!\!\!\!
\tilde{T}_{\!A\;\;\;t}^{{\scriptscriptstyle (1)}\: t}= -4 \pi\, \s
                             \nabla^{2} I(x,y),
                   \;\;\;\;\;\;\;\;\;\;\;\;\;\;
\tilde{T}_{\!A\;\;\;z}^{{\scriptscriptstyle (1)}\: z}= 2 \pi\, \s
                             \nabla^{2} I(x,y),    \nonumber\\
&&\!\!\!\!\!\!\!\!\!\!\!\!\!\!
\tilde{T}_{\!A\;\;\;b}^{{\scriptscriptstyle (1)}\: a}= 2 \pi\, \s
\left(  \d_{ab}
                             \nabla^{2} - \p_{a}\p_{b}\right)I(x,y),
                             \label{quant.s.t.6}
\end{eqnarray}
where $I(x,y)\!\equiv\! \int d^{4}x'\: H(-x',\bar{\mu})\, \d(x+x')
\d(y+y')$. This integral is computed in Appendix A with the result
\begin{eqnarray}
I&=&\frac{1}{2 \pi} \, {\cal P}\!f \left(\frac{1}{r^{2}} \right)+\left(
\ln \frac{\bar{\mu}}{2} +\gamma \right) \d(x)\d(y) \nonumber\\
&=& \lim_{\epsilon \rightarrow 0^{+}} \left\{\frac{1}{2 \pi} \,
\frac{1}{r^{2}}
\, \t(r-\epsilon ) + \left( \ln \frac{\bar{\mu} \epsilon}{2} +\gamma
\right) \d(x)\d(y)
\right\}.  \label{I}
\end{eqnarray}
Using the result given in ref. \c{Schwartz} $\nabla^{2} {\cal P}\!f
\left(r^{-2} \right)\= 4 \, {\cal P}\!f \left(r^{-4} \right)\-
2 \pi \nabla^{2} \left(\d(x)\d(y)\right)$ one can substitute in
these expressions
\be
\nabla^{2}\! I= \frac{2}{\pi}   \,{\cal P}\!f \left(\frac{1}{r^{4}}
\right)+\left(\ln \frac{\bar{\mu}}{2} +\gamma-1 \right) \nabla^{2}\!
\left(\d(x)\d(y)\right).
\ee
These results for
$\tilde{T}_{\!A }^{\,{\scriptscriptstyle (1)}{\mu\nu}}$ agree with the
static components of (\ref{quant.s.t.4}) when $r\!\neq\!0$.

For the time dependent terms
$\tilde{T}_{\!A }^{{\scriptscriptstyle (2)}{\mu\nu}}$
the calculation is considerably more complicated. Thus, instead of
computing these
exactly, we will introduce the following approximations.

%%%%%%%%%%%%%%%%%%%%%%%%%%%%%%%%%%%%%%%%%%%%%%%%%%%%%%%%%%

\subsection{The approximated quantum stress tensor
$\tilde{T}_{\!A }^{{\scriptscriptstyle (2)}{\mu\nu}}$}

%%%%%%%%%%%%%%%%%%%%%%%%%%%%%%%%%%%%%%%%%%%%%%%%%%%%%%%%%%

We introduce here some approximations in
$\tilde{T}_{\!A }^{{\scriptscriptstyle (2)}{\mu\nu}}$. First, instead of
working out the exact expresions for these terms to include $r\=0$, we
will introduce a cut-off radius $r_{0}$ (which can be viewed as the
physical string radius) and assume that the tensor
$\tilde{T}_{\!A }^{{\scriptscriptstyle (2)}{\mu\nu}}$
derived in (\ref{quant.s.t.4}) is only valid for $r\!\geq\!r_{0}$.
We will make some assumptions for the values of this tensor at
$r\!<\!r_{0}$. At the end of the calculations we will take the limit
$r_{0}\!\rightarrow\!0$. Beside this, in the terms of
(\ref{quant.s.t.4}) which correspond to
$\tilde{T}_{\!A }^{{\scriptscriptstyle (2)}{\mu\nu}}$ we will
substitute the distributions $f_{C}(b)$ and $g(b)$ by $2 \, \t(t\-r)$
and zero respectively. That is, we make a sudden approximation assuming
that the values of
$\tilde{T}_{\!A\,\mu\nu}^{\,{\scriptscriptstyle (2)}}$ change
suddenly at the light cone $t\=r$. This seems justified in view of the
fact that such terms, as we have seen before, settle quickly to the
final static values. But now, to ensure the conservation of
$\tilde{T}_{\!A\,\mu\nu}^{\,{\scriptscriptstyle (2)}}$
on the light cone $t\=r$ we need to add terms proportional to
$\d(t\-r)$, which have the same singular behavior on the light cone as
the stress tensor (\ref{quant.s.t.4}). Imposing also that
$\tilde{T}_{\!A\;\;\;\;\;\;\mu}^{\,{\scriptscriptstyle (2)}\,\mu}\=0$
we obtain the following approximated stress tensor for
$r\!\geq\!r_{0}$:

\be
\tilde{T}_{\!A\;\;\;\;\;\;\nu}^{\,{\scriptscriptstyle (2)}\,\mu}=
\frac{2\s\tau}{r^{4}} \left[ \t(t\-r)
   \left(
   \begin{array}{rrrr}
   2 &    &   &   \\
     & -1 &   &   \\
     &    & 3 &   \\
     &    &   &  -4
   \end{array} \right)+ r \,\d(t\-r)
   \left(
   \begin{array}{rrrr}
   1 & -1 &   &   \\
   1 & -1 &   &   \\
     &    & 3 &   \\
     &    &   &  -3
   \end{array} \right)
\right],   \label{approx.s.t.out}
\ee
in polar coordinates. Now we have to make an assumption for the values
of $\tilde{T}_{\!A\,\mu\nu}^{\,{\scriptscriptstyle (2)}}$ at
$r\!<\!r_{0}$. We take a stress tensor with terms proportional to
$\t(t\-r)$ and $r \,\d(t\-r)$ as before, but which is constant for
$t\!>\!r$. Imposing that it is conserved,
$\nabla_{\!\mu} \tilde{T}_{\!A }^{\,{\scriptscriptstyle (2)}{\mu\nu}}
\=0$, and traceless,
$\tilde{T}_{\!A\;\;\;\;\;\;\mu}^{\,{\scriptscriptstyle (2)}\,\mu}\=0$,
we find for $r\!<\!r_{0}$:
\be
\tilde{T}_{\!A\;\;\;\;\;\;\nu}^{\,{\scriptscriptstyle (2)}\,\mu}=
-\frac{2\s\tau}{r_{0}^{4}} \left[ \t(t\-r)
   \left(
   \begin{array}{rrrr}
   2 &    &   &   \\
     & 1 &   &   \\
     &    & 1 &   \\
     &    &   &  -4
   \end{array} \right)+ r \,\d(t\-r)
   \left(
   \begin{array}{rrrr}
   -1 & 1 &   &   \\
   -1 & 1 &   &   \\
     &    & 1 &   \\
     &    &   &  -1
   \end{array} \right)
\right].         \label{approx.s.t.in}
\ee
Note that when we take the divergence of the complete
$\tilde{T}_{\!A\;\;\;\;\;\;\nu}^{\,{\scriptscriptstyle (2)}\,\mu}$ we
have the step functions $\t(r_{0}\-r)$ and $\t(r\-r_{0})$ multiplying
the above expressions which have to be derived too.

It can be seen also that the values of
$\tilde{T}_{\!A\;\;\;\;\;\;\nu}^{\,{\scriptscriptstyle (2)}\,\mu}$
for $r\!<\! r_{0}$ and $t\!>\!r$ obtained in (\ref{approx.s.t.in})
correspond to the values at $r\=0$ of the quantum stress tensor for a
model based on the classical stress tensor for the string given in
refs.\ \c{Gott III 85,His 85}. In such a model the string has a finite
radius $r_{0}$ inside which the classical stress tensor is assumed to
be constant; this tensor is given by
\be
T_{\!c \;\: \nu}^{\,\mu}=\varepsilon \,\t(r_{0}-r)\: {\rm diag}
(1,0,0,\tau), \label{class.s.t.2}
\ee
where we have introduced the $\tau$ parameter in order to allow for the
possibility of a straight wiggly string. The string energy density
$\varepsilon$ is related to its energy per unit length $\mu$ (see refs.
\c{Gott III 85,His 85}) by $4G\mu\=1\-\cos {\displaystyle \left(r_{0}
\sqrt{8 \pi G \varepsilon}\right)}$. At first order in $G\mu$ this
relation gives $\pi G \varepsilon r_{0}^{2}\= G\mu+O(G^{2}\mu^{2})$.
Using the previous expressions (\ref{quant.s.t.})-(\ref{A & B}), we can
calculate the stress tensor $\langle T_{\mu\nu}\rangle$ to first order
in $G\mu$ for a free massless conformally coupled scalar field with the
classical source (\ref{class.s.t.2}). The values of such tensor at
$r\=0$ agree with these of the $\t(t\-r)$ terms in
(\ref{approx.s.t.in}) and this gives further justification to the
approximation taken.

%%%%%%%%%%%%%%%%%%%%%%%%%%%%%%%%%%%%%%%

\section{The back-reaction metric}
\setcounter{equation}{0}

%%%%%%%%%%%%%%%%%%%%%%%%%%%%%%%%%%%%%%%

In this section we compute the correction to the gravitational field
due to the vacuum polarization of matter fields at one-loop order given
in the previous section. For this we solve the semiclassical
correction to Einstein's equations,
\be
G^{\mu\nu}(x)=-8 \pi G\,[T_{\!c}^{\,\mu\nu}(x)+\langle T^{\mu\nu}(x)
\rangle],  \label{s-c eqs.}
\ee
in the linear approximation. Within this approximation the metric
tensor can be written as
\be
g_{\mu\nu}=\eta_{\mu\nu}+h_{\mu\nu}+\tilde{h}_{\mu\nu},
\ee
where $h_{\mu\nu}$ is the metric perturbation due to the classical
string stress tensor $T_{\!c}^{\,\mu\nu}$ and $\tilde{h}_{\mu\nu}$ is
the quantum correction to the metric which is of order $\hbar$. The
classical part, $h_{\mu\nu}$, has already been computed in section 2,
and we note here that $\langle T^{\mu\nu}(x)\rangle$ depends on that
part only, since including $\tilde{h}_{\mu\nu}$ in $\langle
T^{\mu\nu}(x)\rangle$ would lead to terms of order $\hbar^{2}$
which we neglect. Thus eq.\ (\ref{s-c eqs.}) leads in the harmonic
gauge $(\tilde{h}^{\mu\nu}\!\!-\!\frac{1}{2}\,\eta^{\mu\nu}
\tilde{h}^{\alpha}_{\;\:\alpha}),_{\,\nu}=0$ to the equation
\be
\Box \tilde{h}_{\mu\nu}=-16\pi G \tilde{S}_{\mu\nu},  \label{b-r eqs.}
\ee
where $\tilde{S}_{\mu\nu}\equiv \langle T_{\mu\nu}\rangle-\!
\frac{1}{2}\,\eta_{\mu\nu}\langle T^{\alpha}_{\;\:\alpha}\rangle$. A
formal solution of this equation is given as an initial
value problem by (\ref{Kirchhoff}), where one substitutes $h_{\mu\nu}$
by $\tilde{h}_{\mu\nu}$ and $S_{\mu\nu}$ by $\tilde{S}_{\mu\nu}$.
Now from the decomposition (\ref{quant.s.t.5}) for the stress
tensor and the separation of $\tilde{T}_{\!B\,\mu\nu}$
into a static and a time dependent parts,
$\tilde{T}_{\!B\,\mu\nu}^{\,{\scriptscriptstyle (1)}}$ and
$\tilde{T}_{\!B\,\mu\nu}^{\,{\scriptscriptstyle (2)}}$, it is clear
that a similar separation can be performed in $\tilde{S}_{\mu\nu}$,
{\it i.e.}
\be
\tilde{S}_{\mu\nu}\=
\tilde{T}_{\!A\,\mu\nu}^{\,{\scriptscriptstyle (1)}}+
\tilde{S}_{\!B\,\mu\nu}^{\,{\scriptscriptstyle (1)}}+
\tilde{T}_{\!A\,\mu\nu}^{\,{\scriptscriptstyle (2)}}+
\tilde{S}_{\!B\,\mu\nu}^{\,{\scriptscriptstyle (2)}},
\ee
where we use that $\tilde{T}_{\!A\,\mu\nu}$ is traceless.
Then (\ref{b-r eqs.}) leads to likewise separate quantum perturbations:
$\tilde{h}_{\mu\nu}\=
\tilde{h}_{\!A\,\mu\nu}^{\,{\scriptscriptstyle (1)}}+
\tilde{h}_{\!B\,\mu\nu}^{\,{\scriptscriptstyle (1)}}+
\tilde{h}_{\!A\,\mu\nu}^{\,{\scriptscriptstyle (2)}}+
\tilde{h}_{\!B\,\mu\nu}^{\,{\scriptscriptstyle (2)}}\!\equiv \!
\tilde{h}^{{\scriptscriptstyle (1)}}_{\mu\nu}+
\tilde{h}^{{\scriptscriptstyle (2)}}_{\mu\nu}$. The main advantage
of this separation is that the initial conditions at the surface $t\=0$
for $\tilde{h}^{{\scriptscriptstyle (2)}}_{\mu\nu}$ are simply
\be
\tilde{h}^{{\scriptscriptstyle (2)}}_{\mu\nu}{\Big\vert}_{t=0}= \p_{t}
\tilde{h}^{{\scriptscriptstyle (2)}}_{\mu\nu}{\Big\vert}_{t=0}=0,
\ee
consequently, the explicit form of
$\tilde{h}_{\!A\,\mu\nu}^{\,{\scriptscriptstyle (2)}}(t,\vec{x})$ is
given by
\be
\tilde{h}_{\!A\,\mu\nu}^{\,{\scriptscriptstyle (2)}}(t,\vec{x}\,)=-16
\pi G\int_{0}^{t} dt'
\int_{_{\displaystyle I\hspace{-0.7 ex} R^{3}}} \!\!\!\! d^{3}x' \:
D_{R}(x-x') \,
\tilde{T}_{\!A\,\mu\nu}^{\,{\scriptscriptstyle (2)}}(t',\vec{x\,}').
\label{h2}
\ee
Similarly $\tilde{h}_{\!B\,\mu\nu}^{\,{\scriptscriptstyle (2)}}$ is
given by this expression changing
$\tilde{T}_{\!A\,\mu\nu}^{\,{\scriptscriptstyle (2)}}$ by
$\tilde{S}_{\!B\,\mu\nu}^{\,{\scriptscriptstyle (2)}}$.

We are only interested in finding these semiclassical perturbations
$\tilde{h}_{\mu\nu}$ outside the string, that is, for $r\!\neq\!0$.

%%%%%%%%%%%%%%%%%%%%%%%%%%%%%%%%%%

\subsection{The static part}

%%%%%%%%%%%%%%%%%%%%%%%%%%%%%%%%%%

For the static correction
$\tilde{h}^{{\scriptscriptstyle (1)}}_{\mu\nu}$
eq.\ (\ref{b-r eqs.}) reduces to
\be
\nabla^{2}  \tilde{h}^{{\scriptscriptstyle (1)}}_{\mu\nu}
=16\pi G \,\tilde{S}^{{\scriptscriptstyle (1)}}_{\mu\nu},
\label{static eqs}
\ee
where $\tilde{S}^{\,{\scriptscriptstyle (1)}}_{\mu\nu}\!\equiv \!
\tilde{T}_{\!A\,\mu\nu}^{\,{\scriptscriptstyle (1)}}+
\tilde{S}_{\!B\,\mu\nu}^{\,{\scriptscriptstyle (1)}}$.
It is very simple to solve these equations for $r\!\neq\!0$ making use
of the cylindrical symmetry of the problem. Notice that
$\tilde{S}_{\!B\,\mu\nu}^{\,{\scriptscriptstyle (1)}}$
has support on $r\=0$, so from (\ref{quant.s.t.4}), we have in the
polar coordinate basis
$\tilde{S}^{\,{\scriptscriptstyle (1)}\,\nu}_{\;\;\,\mu}\=\tilde{T}
_{\!A\;\mu}^{\,{\scriptscriptstyle (1)}\,\nu}\= - 2 \s\,r^{-4}\:
{\rm diag}(4,1,-3,-2)$ for $r\!\neq\!0$. It is now easy to see that
the following functions (which are defined in terms of the metric
perturbations),
$\tilde{h}^{\,\scriptscriptstyle (1)}_{1}\!\equiv\!
\tilde{h}^{\,{\scriptscriptstyle (1)} t}_{\;\;\;\;\,t}$,
$\tilde{h}^{\,\scriptscriptstyle (1)}_{2}\!\equiv\!
\tilde{h}^{\,{\scriptscriptstyle (1)} z}_{\;\;\;\;\,z}$,
$\tilde{h}^{\,\scriptscriptstyle (1)}_{3}\!\equiv\!
\tilde{h}^{\,{\scriptscriptstyle (1)} r}_{\;\;\;\;\,r}\+
\tilde{h}^{\,{\scriptscriptstyle (1)} \t}_{\;\;\;\;\,\t}$ and
$\tilde{h}^{ \,\scriptscriptstyle (1)}_{4}\!\equiv\!
\tilde{h}^{\, {\scriptscriptstyle (1)} r}_{\;\;\;\;\,r}\-
\tilde{h}^{\,{\scriptscriptstyle (1)} \t}_{\;\;\;\;\,\t}$, depend on
$r$ only and satisfy the equations, for $r\!\neq\!0$,
\be
\left( \frac{\!\! d^{2}}{dr^{2}}+\frac{1}{r} \frac{\!\!d}{dr}
-4 \,\frac{\d_{k4}}{r^{2}} \right)
\tilde{h}^{\scriptscriptstyle (1)}_{k}= \frac{64 \pi\s G}{r^{4}}\,
\gamma_{k},
\ee
where $\gamma_{1}\=\gamma_{4}\=-2$ and $\gamma_{2}\=\gamma_{3}\=1$.
The harmonic gauge condition in terms of these functions can be
written as
$(\tilde{h}^{\scriptscriptstyle (1)}_{1} \+
\tilde{h}^{\scriptscriptstyle (1)}_{2}\-
\tilde{h}^{\scriptscriptstyle (1)}_{4} )_{,r}\=
2  r^{-1} \tilde{h}^{\scriptscriptstyle (1)}_{4}$.
The general solution of these differential equations can be
expressed as a linear combination of terms such as $r^{2}$, $\ln r$,
$r^{-2}$ and $r^{-2} \ln r$. Finally, after a gauge
transformation, within the harmonic gauge, the solution can be written
as
\be
\begin{array}{ll}
\tilde{h}^{\scriptscriptstyle (1)}_{1}={\displaystyle -32 \pi \s G \,
\frac{1}{r^{2}} + \s G \, A
                                           \,\ln\! \left(
\frac{r}{R_{1}} \right)},    &
\;\;\tilde{h}^{\scriptscriptstyle (1)}_{2}={\displaystyle 16 \pi \s G
\, \frac{1}{r^{2}} - \s G \, A
                                           \,\ln\! \left(
\frac{r}{R_{2}} \right)}, \\
\tilde{h}^{\scriptscriptstyle (1)}_{3}={\displaystyle 16 \pi \s G \,
\frac{1}{r^{2}}},                          &
\;\;\tilde{h}^{\scriptscriptstyle (1)}_{4}={\displaystyle 32 \pi \s G
\, \frac{1}{r^{2}} \,\ln\! \left( \frac{r}
                                           {R} \right)},
\label{s.quant.metric}
\end{array}
\ee
where the constants of integration $R_{1}$, $R_{2}$ and $R$ have
dimensions of length and $A$ has dimensions of $({\rm length})^{-2}$.
Notice that there is a gauge freedom within the harmonic gauge to set
the constants $R_{1}$, $R_{2}$ and $R$ to a fixed value, {\it i.e.}\ a c
hange in the values of these constants is equivalent to an harmonic
gauge transformation. The values of these constants in the solution
for $r\!\neq\!0$ can be determined if we solve the equations
(\ref{static eqs}) including $r\=0$. However, without working out the
explicit solution, we can use dimensional arguments to see that $A$
must be zero. Note that the only dimensional constant parameter which
can be used to make $\tilde{h}^{{\scriptscriptstyle (1)}}_{\mu\nu}/\s
G$ is $\bar{\mu}$, but since the dependence in this parameter must be
logarithmic we are forced to take $A\=0$, on the other hand $R$
will be proportional to $1/\bar{\mu}$. In fact, we can set
$R\=1/\bar{\mu}$ after an harmonic gauge transformation for
$r\!\neq\!0$.

By doing a little more work, we can also arrive at the above solutions
solving explicitly equations (\ref{static eqs}) with the inclusion of
$r=0$. In fact, from (\ref{quant.s.t.B}) and (\ref{quant.s.t.6})
it is easy to see that the equations for
$\tilde{h}^{\scriptscriptstyle (1)}_{1}$,
$\tilde{h}^{\scriptscriptstyle (1)}_{2}$
and $\tilde{h}^{{\scriptscriptstyle (1)} x}_{\;\;\;\;x}\+
\tilde{h}^{{\scriptscriptstyle (1)} y}_{\;\;\;\;y}$ are
\begin{eqnarray}
&&{\nabla}^{2} \left[\tilde{h}^{\scriptscriptstyle (1)}_{1}+
    16 \pi^2 \s G \left( 4 \, I- \frac{1}{6} \,\d(x) \d(y) \right)
    \right]=0, \nonumber\\
&&{\nabla}^{2} \left[\tilde{h}^{\scriptscriptstyle (1)}_{2}-
    16 \pi^2 \s G \left( 2 \, I+ \frac{1}{6} \,\d(x) \d(y) \right)
    \right]=0,
    \nonumber\\
&&{\nabla}^{2}
    \left[\tilde{h}^{{\scriptscriptstyle (1)} x}_{\;\;\;\;x}\+
\tilde{h}^{{\scriptscriptstyle (1)} y}_{\;\;\;\;y}
        -32 \pi^2 \s G  \left( I+ \frac{1}{3} \,\d(x) \d(y)
\right)\right]=0,
\end{eqnarray}
where $I$ is given by (\ref{I}). These give the solutions
$\tilde{h}^{\scriptscriptstyle (1)}_{1}
\= -16 \pi^2 \s G \left( 4\, I-\frac{1}{6} \, \d(x) \d(y) \right)$,
$\tilde{h}^{\scriptscriptstyle (1)}_{2}\= 16 \pi^2 \s G
\left( 2\, I+\frac{1}{6} \, \d(x) \d(y) \right)$ and
$\tilde{h}^{{\scriptscriptstyle (1)} x}_{\;\;\;\;x}\+
\tilde{h}^{{\scriptscriptstyle (1)} y}_{\;\;\;\;y}\= 32 \pi^2 \s G
\left( I+\frac{1}{3} \, \d(x) \d(y) \right)$, which
for $r\!\neq\!0$ reduce to (\ref{s.quant.metric}) with $A\=0$. If we
write the remaining two equations for
$\tilde{h}^{{\scriptscriptstyle (1)} x}_{\;\;\;\;x}\-
\tilde{h}^{{\scriptscriptstyle (1)} y}_{\;\;\;\;y}$ and
$\tilde{h}^{{\scriptscriptstyle (1)} x}_{\;\;\;\;y}$, using that
$\d(x)\d(y)\= \frac{1}{2 \pi} \, {\nabla}^{2} \ln r$,
we see that for $r\!\neq\!0$ one has
$\tilde{h}^{\scriptscriptstyle (1)}_{4}\= 32 \pi \s G \,
r^{-2} \,\ln\! \left(\kappa \bar{\mu} r \right)$, where $\kappa$ is some
numerical factor. So we finally have for $r\!\neq\!0$
\be
\begin{array}{ll}
\tilde{h}^{\scriptscriptstyle (1)}_{1}={\displaystyle -32 \pi \s G \,
\frac{1}{r^{2}},}    &
\;\;\;\;\;\;\tilde{h}^{\scriptscriptstyle (1)}_{2}={\displaystyle 16
\pi \s G \, \frac{1}{r^{2}},} \\
\tilde{h}^{\scriptscriptstyle (1)}_{3}={\displaystyle 16 \pi \s G \,
\frac{1}{r^{2}}},  &
\;\;\;\;\;\;\tilde{h}^{\scriptscriptstyle (1)}_{4}={\displaystyle 32
\pi \s G \, \frac{1}{r^{2}} \,\ln \bar{\mu} r}.
                                               \label{s.quant.metric 2}
\end{array}
\ee

This solution can also be obtained by the procedure which will use next
to obtain the time dependent metric perturbations. Namely, we
introduce a cut-off radius $r_{0}$ and take the limit
$r_{0}\!\rightarrow\!0$ at the end of the calculation. For this we use
an approximated stress tensor
$\tilde{T}_{\!A\,\mu\nu}^{\,{\scriptscriptstyle (1)}}$
for $r\!<\! r_{0}$ as in section 3.3.

Let us now consider the contribution
$\triangle \tilde{\t}^{\scriptscriptstyle (1)}$
to the deficit angle of the two-surfaces $t\={\rm const.}$,
$z\={\rm const.}$ due
to these static corrections to the metric. Substituting in
(\ref{def angle}) the values of
$\tilde{h}^{{\scriptscriptstyle (1)}}_{\mu\nu}$
given by (\ref{s.quant.metric 2}) we find
\be
\triangle \tilde{\t}^{\scriptscriptstyle (1)}=
\frac{32 \pi^{2} \s G}{r^{2}}.
\label{def angle 1}
\ee

We can now calculate the static quantum correction to the Riemann
tensor. Using the expressions derived in Appendix B we find the
following non-vanishing components for $r\!\neq\!0$
\begin{eqnarray}
&&\hspace*{-10ex}\tilde{R}^{\scriptscriptstyle (1)}_{\,xzxz}=
\frac{16 \pi \s G}{r^{4}}\,
            (2\cos 2\t+1), \;\;\;\;\;
 \tilde{R}^{\scriptscriptstyle (1)}_{\,xzyz}=
             \frac{32 \pi \s G}{r^{4}}\, \sin 2\t, \;\;\;\;\;
\tilde{R}^{\scriptscriptstyle (1)}_{\,xyxy}=\frac{32 \pi \s G}{r^{4}},
\nonumber\\
&&\hspace*{-10ex}\tilde{R}^{\scriptscriptstyle (1)}_{\,txtx}\,=
             \frac{32 \pi \s G}{r^{4}}\, (2\cos 2\t+1),  \;\;\;\;\;
 \tilde{R}^{\scriptscriptstyle (1)}_{\,txty}\,=
             \frac{64 \pi \s G}{r^{4}}\, \sin 2\t,
\label{s.Riemann}
\end{eqnarray}
and also $\tilde{R}^{\scriptscriptstyle (1)}_{\,yzyz}$ and
$\tilde{R}^{\scriptscriptstyle (1)}_{\,tyty}$, which are obtained from
$\tilde{R}^{\scriptscriptstyle (1)}_{\,xzxz}$ and
$\tilde{R}^{\scriptscriptstyle (1)}_{\,txtx}$
interchanging $\cos \t$ by $\sin \t$.

%%%%%%%%%%%%%%%%%%%%%%%%%%%%%%%%%%%%%%

\subsection{The time dependent part}

%%%%%%%%%%%%%%%%%%%%%%%%%%%%%%%%%%%%%%

Let us first consider the contribution to the quantum perturbations to
the metric $\tilde{h}_{\!A\,\mu\nu}^{\,{\scriptscriptstyle (2)}}$
coming from $\tilde{T}_{\!A\,\mu\nu}^{\,{\scriptscriptstyle (2)}}$
given by (\ref{approx.s.t.out}) and (\ref{approx.s.t.in}).
This solution is given by (\ref{h2}).
Defining, as we have done before,
$\tilde{h}^{\,{\scriptscriptstyle (2)}}_{\!A\,_{\scriptstyle 1}}\!
\equiv\!
\tilde{h}^{{\scriptscriptstyle (2)} t}_{\!A\;\;t}$,
$\tilde{h}^{\,{\scriptscriptstyle (2)}}_{\!A\,_{\scriptstyle 2}}\!
\equiv\!
\tilde{h}^{{\scriptscriptstyle (2)} z}
_{\!A\;\;z}$,
$\tilde{h}^{\,{\scriptscriptstyle (2)}}_{\!A\,_{\scriptstyle 3}}\!
\equiv\!
\tilde{h}^{{\scriptscriptstyle (2)} r}
_{\!A\;\;r}\+\tilde{h}^{{\scriptscriptstyle (2)} \t}
_{\!A\;\;\t}$,
$\tilde{h}^{\,{\scriptscriptstyle (2)}}_{\!A\,_{\scriptstyle 4}}\!
\equiv\!
\tilde{h}^{{\scriptscriptstyle (2)} r}
_{\!A\;\;r}\-\tilde{h}^{{\scriptscriptstyle (2)} \t}_{\!A\;\;\t}$ and
$\tilde{h}^{\,{\scriptscriptstyle (2)}}_{\!A\,_{\scriptstyle 5}}\!
\equiv\!
\tilde{h}^{{\scriptscriptstyle (2)} r}_{\!A\;\;t}$,
this leads for $(t\-r)\!>\!2r_{0}$ and $r\!>\!r_{0}$ to the solution
\begin{eqnarray}
&&\hspace*{-10ex}
\tilde{h}^{\,{\scriptscriptstyle (2)}}_{\!A\,_{\scriptstyle k}}=
-\frac{32 \s \tau G}{r^{2}}
\int_{0}^{\pi} d\t \: g_{k}(\t)
\;\,\Biggl\{\frac{1}{x_{0}^{4}} \int_{0}^{x_{0}} dx \: x \:
\left[ C_{k}+\bar{C}_{k} \: x\, \frac{\p}{\p b} \right]  \nonumber\\
&&\hspace*{20ex}+ \int_{x_{0}}^{\frac{1}{2} \left(
\frac{b^2-1}{b-\cos \t} \right)} \frac{dx}{x^{3}} \:
\left[ D_{k}+\bar{D}_{k}
\: x\, \frac{\p}{\p b} \right] \Biggr\}\; f(b;x,\t),
\label{t.d.quant.metric}
\end{eqnarray}
where $b\!\equiv\!t/r$, $x_{0}\!\equiv\!r_{0}/r$ and
\begin{eqnarray}
&&\hspace*{-6ex} C_{k} \equiv -2 \,\d_{k1}+4 \,\d_{k2}-2 \,\d_{k3},
   \hspace{10ex}  \bar{C} _{k} \equiv \d_{k1}+\d_{k2}-2 \,\d_{k3}+
\d_{k5}, \nonumber\\
&&\hspace*{-6ex} D_{k} \equiv 2 \,\d_{k1}-4 \,\d_{k2}+2 \,\d_{k3}-4 \,
\d_{k4},
   \hspace{4ex}  \bar{D} _{k} \equiv \d_{k1}-3  \,\d_{k2}+2 \,\d_{k3}-
4 \,\d_{k4}
                 +\d_{k5},      \nonumber\\
&&\hspace*{-6ex} g_{k}(\t) \equiv \left\{ \begin{array}{ll}
                                      1 ,       & k=1,2,3    \\
                                      \cos 2\t, & k=4     \\
                                      \cos \t , & k=5,
                                   \end{array}
                           \right.       \nonumber\\
&&\hspace*{-6ex} f(b;x,\t) \equiv  \ln \left[b-x+ \sqrt{2x (\cos \t-b)+
b^{2}-1} \right]
      -\frac{1}{2} \, \ln (x^{2}-2x \cos \t+1).
\end{eqnarray}

We now expand the solution (\ref{t.d.quant.metric}) in terms of
$x_{0}\!\equiv\!r_{0}/r$ neglecting terms of order $x_{0}$ wich will
vanish in the limit $r_{0}\!\rightarrow\!0$. We find
\be
\tilde{h}^{\,{\scriptscriptstyle (2)}}_{\!A\,_{\scriptstyle k}}
=-\frac{32 \pi \s \tau G}{r^{2}}\,
\left[ \ln x_{0}\, \Phi_{k}^{\rm \scriptscriptstyle II}(b)+
\Phi_{k}^{\rm \scriptscriptstyle I}(b)+O(x_{0}) \right],
\label{metric 2}
\ee
where
\begin{eqnarray}
&&\hspace*{-6ex}\Phi_{1}^{\rm \scriptscriptstyle II}=- \Phi_{3}^{\rm
\scriptscriptstyle II} \equiv
   \frac{1}{2} \, \frac{b}{ (b^2-1)^{3/2}}, \;\;\;\;\;\;\;\;
   \Phi_{2}^{\rm \scriptscriptstyle II}\equiv 0, \;\;\;\;\;\;\;\;
   \Phi_{4}^{\rm \scriptscriptstyle II}\equiv\frac{1}{2} \,
\frac{b}{ (b^2-1)^{3/2}}
   \, (2b^2-3),       \nonumber\\
&&\hspace*{-6ex}\Phi_{5}^{\rm \scriptscriptstyle II}\equiv \frac{1}{2}
\, \frac{1}{ (b^2-1)^{3/2}},       \nonumber\\
&&\hspace*{-6ex}\Phi_{1}^{\rm \scriptscriptstyle I}\equiv  \frac{1}{2}
   \left\{ -1-\frac{1}{ (b^2-1) }+ \frac{b}{ (b^2-1)^{3/2}}\, \left[
\frac{9}{4}+
   {\rm arccosh }\, b- \ln 4 (b^2-1) \right] \right\},       \nonumber\\
&&\hspace*{-6ex}\Phi_{2}^{\rm \scriptscriptstyle I}\equiv 1+\frac{1}{2}
 \, \frac{b}{ (b^2-1)^{3/2}},        \nonumber\\
&&\hspace*{-6ex}\Phi_{3}^{\rm \scriptscriptstyle I}\equiv \frac{1}{2}
   \left\{ -1+\frac{1}{ (b^2-1) }- \frac{b}{ (b^2-1)^{3/2}}\, \left[
\frac{13}{4}+
   {\rm arccosh }\, b- \ln 4 (b^2-1) \right] \right\},       \nonumber\\
&&\hspace*{-6ex}\Phi_{4}^{\rm \scriptscriptstyle I}\equiv 3-
\frac{(2b^6-2b^2-3)}{(b^2-1)}+
   \frac{1}{4}\, \frac{b}{ (b^2-1)^{3/2}} \, (2b^2-1)\, (4b^4-11)
   +{\rm arccosh }\, b       \nonumber\\
&&\!+\frac{1}{2} \, \frac{b}{ (b^2-1)^{3/2}} \, (2b^2-3)\, \biggl[3
\ln 2b-
   2 \: {\rm arccosh }\, b-\ln 4 (b^2-1)        \nonumber\\
&&\hspace*{25ex}  -\frac{5}{4}\,\frac{1}{b^{2}}\,
   \mbox{}_{3}\hspace*{-0.1ex}F_{2}\! \left(1,1,\frac{7}{2};2,4;
\frac{1}{b^{2}}\right)
   \biggr],         \nonumber\\
&&\hspace*{-6ex}\Phi_{5}^{\rm \scriptscriptstyle I}\equiv  \frac{1}{2}
   \left\{-\frac{b}{ (b^2-1) }+ \frac{1}{ (b^2-1)^{3/2}}\, \left[
\frac{5}{4}+b^2+
   {\rm arccosh }\, b- \ln 4 (b^2-1) \right] \right\}.
                                                  \label{functions}
\end{eqnarray}
It is easy to see that we can change in (\ref{metric 2}) $\ln x_{0}$
by $\ln(R/r)$, where $R$ is an arbitrary constant with dimensions of
length. For this we make the following harmonic gauge transformation
\begin{eqnarray}
&&\hspace*{-8ex}\xi_{t}=-8 \pi \s \tau G \, \frac{1}{r \sqrt{b^{2}-1}}
\, \ln \frac{r_{0}}{R} , \;\;\;\;\;\;\;\;\;
  \xi_{z}=0,         \nonumber\\
&&\hspace*{-8ex}\xi_{x}=8 \pi \s \tau G \,
\frac{b \cos\t}{r\sqrt{b^{2}-1}}\,
  \ln \frac{r_{0}}{R} , \;\;\;\;\;\;\;\;\;\;\;
 \xi_{y}=8 \pi \s \tau G \,\frac{b \sin\t}{r\sqrt{b^{2}-1}}\,
  \ln \frac{r_{0}}{R}.
\label{gauge}
\end{eqnarray}
We can also set $R\=1/\bar{\mu}$ and,
after this, we can take the limit $r_{0}\!\rightarrow\!0$ in
$\tilde{h}^{\,{\scriptscriptstyle (2)}}_{\!A\,_{\scriptstyle k}}$.

Let us now consider the terms
$\tilde{h}_{\!B\,\mu\nu}^{\,{\scriptscriptstyle (2)}}$.
Substituting $\tilde{T}_{\!A\,\mu\nu}^{\,{\scriptscriptstyle (2)}}$ by
$\tilde{S}_{\!B\,\mu\nu}^{\,{\scriptscriptstyle (2)}}$ in (\ref{h2})
and defining $\tilde{h}^{\,{\scriptscriptstyle (2)}}_{\!B\,_{
\scriptstyle k}}$ as before we find for $r\!\neq\!0$
\be
\tilde{h}^{\,{\scriptscriptstyle (2)}}_{\!B\,_{\scriptstyle k}}
=-\frac{32 \pi \s \tau G}{r^{2}}\; \frac{1}{6} \: {\cal P}\!f \left[
\Phi_{k}^{\rm \scriptscriptstyle II}(b) \, \t(b-1) \right],
\ee
where the functions $\Phi_{k}^{\rm \scriptscriptstyle II}(b)$ are
defined in (\ref{functions}). It is easy to see that these terms can
be eliminated with a gauge transformation of type (\ref{gauge}). Thus
we finally have for $r\!\neq\!0$ and $b\!>\! 1$
\be
\tilde{h}^{\,{\scriptscriptstyle (2)}}_{\,k}
=-\frac{32 \pi \s \tau G}{r^{2}}\,\left[ \Phi_{k}^{\rm
\scriptscriptstyle I}(b)-
\ln \bar{\mu} r \, \Phi_{k}^{\rm \scriptscriptstyle II}(b) \right].
\label{t.d.quant.metric 2}
\ee

We can now compute the time dependent quantum contribution
$\triangle \tilde{\t}^{\scriptscriptstyle (2)}$ to the deficit angle of
the two-surfaces $t\={\rm const.}$, $z\={\rm const.}$ Substituting in
(\ref{def angle}) the previous values of
$\tilde{h}^{{\scriptscriptstyle (2)}}_{\mu\nu}$ we get
\begin{eqnarray}
&&\hspace*{-6ex}\triangle \tilde{\t}^{\scriptscriptstyle (2)}=
   \frac{8 \pi^{2} \s \tau G}{r^{2}}
   \,\Biggl\{ 2-\frac{1}{(b^2-1)^{2}}\, (4b^4+13b^2-2)
   +\frac{3\,b}{(b^2-1)^{5/2}}\, \biggl[\frac{1}{4}\,(8b^4+8b^2-15)
\nonumber\\
&&\hspace*{17ex}+3 \, \left(\ln 2b-{\rm arccosh }\, b \right)
   -\frac{5}{4}\,\frac{1}{b^{2}}\, \mbox{}_{3}\hspace*{-0.1ex}F_{2}\!
   \left(1,1,\frac{7}{2};2,4;\frac{1}{b^{2}}\right) \biggr]\Biggr\}.
\end{eqnarray}
Expanding this expression in $t^{-1}$ with $r$ fixed one finds
\be
\triangle \tilde{\t}^{\scriptscriptstyle (2)}= \frac{16 \pi^{2} \s
\tau G}{r^{2}}
\left(2-\frac{3}{b^{4}}+O \!\left(\frac{1}{b^{6}}\right)  \right),
\ee
that is, $\triangle \tilde{\t}^{\scriptscriptstyle (2)}$ reaches its
static value $32 \pi^{2} \s \tau G \, r^{-2}$ very quickly. Note that
this contribution due to the time dependent terms is exactly the same
as the contribution $\triangle \tilde{\t}^{\scriptscriptstyle (1)}$ of
(\ref{def angle 1}) when the string is unperturbed ($\tau\=1$).

Finally we can calculate the time dependent part of the Riemann tensor
$\tilde{R}^{\scriptscriptstyle (2)}_{\,\mu\nu\a\b}$. The substitution
of the terms $h_{k}$ of Appendix B by the terms
$\tilde{h}^{\scriptscriptstyle (2)}_{k}$ of
(\ref{t.d.quant.metric 2}) gives the following result for $r\!\neq\!0$
and $b\!>\! 1$
\begin{eqnarray}
&&\hspace*{-4ex}\tilde{R}^{\scriptscriptstyle (2)}_{\,tztz}=-
\frac{24 \pi \s \tau G}{r^{4}}\,
                      \frac{b}{(b^2-1)^{7/2}}\, (2b^2+3),
                                              \nonumber\\
&&\hspace*{-3.5ex}\tilde{R}^{\scriptscriptstyle (2)}_{\,xzxz}=-
\frac{4 \pi \s \tau G}{r^{4}}\,
                      \left[8+\frac{3\,b}{(b^2-1)^{7/2}}\, (2b^2+3)
   +\cos 2\t\, \left(16+\frac{15\, b}{(b^2-1)^{7/2}} \right) \right],
                                              \nonumber\\
&&\hspace*{-3.5ex}\tilde{R}^{\scriptscriptstyle (2)}_{\,xzyz}=-
\frac{4 \pi \s \tau G}{r^{4}}\,
                             \sin 2\t\, \left(16+
                                 \frac{15\, b}{(b^2-1)^{7/2}} \right),
                                              \nonumber\\
&&\hspace*{-3.5ex}\tilde{R}^{\scriptscriptstyle (2)}_{\,txtx}=-
\frac{4 \pi \s \tau G}{r^{4}}\,
                   \Biggl\{ 4-\frac{3\,b}{(b^2-1)^{7/2}}\, (2b^2+3)+
         \cos 2\t \,\Biggl[ 8+\frac{15\,b^{2}}{(b^2-1)^{3}}\, (2b^2+3)
                                              \nonumber\\
&&\hspace*{25ex}-\frac{15\,b}{(b^2-1)^{7/2}}\, \biggl[\frac{1}{4}\,
        (8b^4+8b^2-11)+3 \, \left(\ln 2b-{\rm arccosh }\, b \right)
                                              \nonumber\\
&&\hspace*{39ex}-\frac{5}{4}\,\frac{1}{b^{2}}\,
                                              \mbox{}_{3}
\hspace*{-0.1ex}F_{2}\!
                  \left(1,1,\frac{7}{2};2,4;\frac{1}{b^{2}}\right)
                                              \biggr]\Biggr]\Biggr\},
                                              \nonumber\\
&&\hspace*{-3.5ex}\tilde{R}^{\scriptscriptstyle (2)}_{\,txty}=-
\frac{4 \pi \s \tau G}{r^{4}}\,
        \sin 2\t \,\Biggl\{ 8+\frac{15\,b^{2}}{(b^2-1)^{3}}\, (2b^2+3)
      -\frac{15\,b}{(b^2-1)^{7/2}}\, \biggl[\frac{1}{4}\,(8b^4+8b^2-11)
             \nonumber\\
&&\hspace*{25ex}+3 \, \left(\ln 2b-{\rm arccosh }\, b \right)
                           -\frac{5}{4}\,\frac{1}{b^{2}}\,
                                 \mbox{}_{3}\hspace*{-0.1ex}F_{2}\!
                     \left(1,1,\frac{7}{2};2,4;\frac{1}{b^{2}}\right)
         \biggr]\Biggr\},
                                              \nonumber\\
&&\hspace*{-3.5ex}\tilde{R}^{\scriptscriptstyle (2)}_{\,tzxz}=
\frac{24 \pi \s \tau G}{r^{4}}\,
                      \cos\t \, \frac{1}{(b^2-1)^{7/2}}\, (4b^2+1),
                                              \nonumber\\
&&\hspace*{-3.5ex}\tilde{R}^{\scriptscriptstyle (2)}_{\,txxy}=
\frac{12 \pi \s \tau G}{r^{4}}\,
             \sin \t \,\Biggl\{ \frac{b}{(b^2-1)^{3}}\, (8b^4+14b^2+3)
   -\frac{(4b^2+1)}{(b^2-1)^{7/2}}\, \biggl[\frac{1}{4}\,(8b^4+8b^2-15)
              \nonumber\\
&&\hspace*{25ex}+3 \, \left(\ln 2b-{\rm arccosh }\, b \right)
                             -\frac{5}{4}\,\frac{1}{b^{2}}\,
                          \mbox{}_{3}\hspace*{-0.1ex}F_{2}\!
                   \left(1,1,\frac{7}{2};2,4;\frac{1}{b^{2}}\right)
               \biggr]\Biggr\},
                        \nonumber\\
&&\hspace*{-3.5ex}\tilde{R}^{\scriptscriptstyle (2)}_{\,xyxy}=
-\frac{4 \pi \s \tau G}{r^{4}}\,
             \Biggl\{ -4-\frac{1}{(b^2-1)^{3}}\, (16b^6+24b^4+39b^2-4)
                                              \nonumber\\
&&\hspace*{17ex}+\frac{3\,b}{(b^2-1)^{7/2}}\,(2b^2+3)\,
\biggl[\frac{1}{4}\,(8b^4+8b^2-15)
                       +3 \, \left(\ln 2b-{\rm arccosh }\, b \right)
                                              \nonumber\\
&&\hspace*{40ex}-\frac{5}{4}\,\frac{1}{b^{2}}\,
                        \mbox{}_{3}\hspace*{-0.1ex}F_{2}\!
                     \left(1,1,\frac{7}{2};2,4;\frac{1}{b^{2}}\right)
                                              \biggr]\Biggr\},
\label{t.d.Riemann}
\end{eqnarray}
where $\mbox{}_{3}\hspace*{-0.1ex}F_{2}$ denotes a generalized
hypergeometric function. As before
$\displaystyle{\tilde{R}^{\scriptscriptstyle (2)}_{\,yzyz}}$,
$\displaystyle{\tilde{R}^{\scriptscriptstyle (2)}_{\,tyty}}$,
$\displaystyle{\tilde{R}^{\scriptscriptstyle (2)}_{\,tzyz}}$ and
$\displaystyle{\tilde{R}^{\scriptscriptstyle (2)}_{\,tyyx}}$
are obtained interchanging $\cos \t$ by $\sin \t$ in
$\displaystyle{\tilde{R}^{\scriptscriptstyle (2)}_{\,xzxz}}$,
$\displaystyle{\tilde{R}^{\scriptscriptstyle (2)}_{\,txtx}}$,
$\displaystyle{\tilde{R}^{\scriptscriptstyle (2)}_{\,tzxz}}$ and
$\displaystyle{\tilde{R}^{\scriptscriptstyle (2)}_{\,txxy}}$
respectively. Notice that
$\tilde{R}^{\scriptscriptstyle (2)}_{\,\mu\nu\a\b}$
does not depend on the arbitrary parameter $\bar{\mu}$. This is due to
the fact that, as we have seen before, the value of $\bar{\mu}$ can
be changed by a gauge transformation. The static limit of these
components, {\it i.e.}\
${\displaystyle \lim_{t \rightarrow \infty}}
\tilde{R}^{\scriptscriptstyle (2)}_{\,\mu\nu\a\b}
\equiv R^{\scriptscriptstyle L}_{\,\mu\nu\a\b}$, is
\begin{eqnarray}
&&\hspace*{-3.5ex}R^{\scriptscriptstyle L}_{\,xzxz}=
           -\frac{32 \pi \s \tau G}{r^{4}}\, (2\cos 2\t+1), \;\;\;\;
 R^{\scriptscriptstyle L}_{\,xzyz}=-\frac{64 \pi \s \tau G}{r^{4}}\,
            \sin 2\t, \;\;\;\;
 R^{\scriptscriptstyle L}_{\,xyxy}= \frac{32 \pi \s \tau G}{r^{4}},
\nonumber\\
&&\hspace*{-3.5ex}R^{\scriptscriptstyle L}_{\,txtx}\,=-\frac{16 \pi \s
\tau G}{r^{4}}\,
            (2\cos 2\t+1),\;\;\;\;
 R^{\scriptscriptstyle L}_{\,txty}\,=-\frac{32 \pi \s \tau G}{r^{4}}\,
\sin 2\t,
\label{L.Riemann}
\end{eqnarray}
and $R^{\scriptscriptstyle L}_{\,yzyz}$ and
$R^{\scriptscriptstyle L}_{\,tyty}$
are found interchanging $\cos \t$ by $\sin \t$ in
$\tilde{R}^{\scriptscriptstyle L}_{\,xzxz}$ and
$\tilde{R}^{\scriptscriptstyle L}_{\,txtx}$
respectively. From (\ref{t.d.Riemann}) we see that these final static
values are quickly reached since the corrections for large times go at
least as $b^{-4}\=r^{4}/t^{4}$. The final semiclassical Riemann
components are obtained by adding to the classical values
(\ref{class.Riemann}) the back-reaction corrections (\ref{s.Riemann})
and (\ref{t.d.Riemann}).

%%%%%%%%%%%%%%%%%%%%%%%%%%%%%

\section{Conclusions}

%%%%%%%%%%%%%%%%%%%%%%%%%%%%%

In this paper we have derived the back reaction, due to quantum
fluctuations of matter fields, on the gravitational field of a cosmic
string during and after its formation. As matter fields we have just
considered a massless conformally coupled scalar field but the results
are easily extrapolated to $N$ of such fields. As a model of cosmic
string formation we take an initial thin straight rod whose tension
grows suddenly (step approximation) from zero to a maximum value which
corresponds to the mass per unit length if the string is unperturbed,
or to a smaller tension if the string is wiggly.

Within the linear approximation of Einstein's equations we have first
computed the metric perturbations and the corresponding curvature
tensor for both unperturbed and wiggly strings. If the final string is
unperturbed the Newtonian potential vanishes and the spacetime
becomes flat with a deficit angle, but if the string is wiggly there
is a Newtonian force per unit mass which goes like
$G \mu (1\-\tau) r^{-1}$. This force may play an important role in
the formation of wakes behind long strings \c{Vach & Vilen 91}.

We have then computed the vacuum expectation value of the stress tensor
of matter coupled linearly to the string gravitational field. We have
seen that after formation the stress tensor settles quickly to that of
the final unperturbed or wiggly static string. The stress tensor has
an energy density which goes like $N \hbar G \mu r^{-4}$ (as typical
of Casimir type energies), where we have assumed $N$ matter fields
which correspond to massless fields or with masses less than $m_{H}$,
where  $m_{H}$ is the Higgs mass responsible for symmetry breaking.

With such a stress tensor as a source we have computed the
perturbation induced on the gravitational field. We have also computed
the Riemann tensor components, which give essentially the tidal forces,
corresponding to such quantum corrections. The static part of this
tensor correspond to a Newtonian force per unit mass which goes like
$N \hbar G^{2} \mu r^{-3}$. The time dependent part of the
semiclassical perturbations to the metric and the Riemann
tensor, on the other hand, are quite complicated. But, for a given
radius, after a short time $t\! >\! r $ the curvature tensor becomes
static. Unlike for the classical part the quantum correction to the
gravitational field does not differ substantially if the final string
is unperturbed or wiggly.

Let us now discuss the importance of the quantum corrections in the
case of a wiggly string and the unperturbed string after their
formation. In the case of a wiggly string, {\it i.e.}\ $\tau \!<\!1$
(typically $\tau\!\sim\!0.5$) \c{Vach & Vilen 91}, the ratio between
the quantum, $F_{q}$ and classical, $F_{c}$ forces on surrounding
nonrelativistic particles can be estimated in the static limit as
$F_{q}/F_{c}\!\sim\! 64\pi\a \, N \hbar G \,(2\-\tau)/ \left[(1\-\tau)
\, r^{2} \right]$, {\it i.e.}, if $l_{p}$ is the
Planck length,
$F_{q}/F_{c}\!\sim\! 10^{-2} N \left( l_{p}/r \right)^{2}$.
This means that, unless $N$ is unreasonably large, the quantum effects
are always negligible if $r\!>\!\!\!\!\! $\raisebox {-1.1ex}{$\sim $}
$\! 10 l_{p}$. Note that for a cosmic string the smallest value that
$r$ can take is $r_{0}\!\sim\! \hbar/m_{H}$ and, since
$\mu\!\sim\! m_{H}^{2}/ \hbar$, we have
$F_{q}/F_{c}(r)< F_{q}/F_{c} \,(r_{0})\!\sim\! 10^{-2} N G \mu$
which for GUT strings is of order $10^{-5}-10^{-6}$.

For an unperturbed string $\tau\=1$ there is no classical force $F_{c}$.
Thus the quantum correction will be responsible for a Newtonian force
on the classical matter surrounding the string. However, that force
decrases like $r^{-3}$ which means that it becomes negligible very
quickly at macroscopic distances from the string. In this case it is
better to consider the deficit angle which is an important physical
observable. The deficit angle can be written as a sum of a classical
term plus a term of quantum origin $\triangle\t \+\triangle
\tilde{\t}$, using the notation of the last section. As we have seen,
for the classical part $\triangle\t \!\sim\!G \mu$ whereas
for the quantum part
$\triangle\tilde{\t}\!\sim\! 10^{-2} N \hbar G^{2} \mu r^{-2}$,
so that $\triangle \tilde{\t}/\triangle \t \!\sim\!  10^{-2}N
\left( l_{p}/r\right)^{2}$ which is of the same order as the ratio
between the quantum and classical forces given before. The ratio of
the transverse velocities on the string wakes due to the quantum
effect and the classical deficit angle are also of this order. These
results hint that whereas one should not expect any quantum effect at
macroscopic distances on surrounding matter, quantum effects might
start to become important near the string at microscopic distances. In
particular, when two strings cross, there might be a correction,
perhaps in the form of a Casimir like force, due to quantum effects.
However in this case our classical picture breaks down and one must
consider the dynamics of the Higgs fields themselves at microscopical
level.

%%%%%%%%%%%%%%%%%%%%%%%%%%%%%%%%

\section{Acknowledgements}

%%%%%%%%%%%%%%%%%%%%%%%%%%%%%%%%

We are grateful to Jaume Garriga and Carlos Lousto for helpful
discussions. This work has been partially supported by a CICYT Research
Project number \mbox{AEN93-0474}.

\newpage

%%%%%%%%%%%%%%%%%%%%%%%%%%%%%%%%%%

{\noindent \Large \bf Appendix A}
\appendix

%%%%%%%%%%%%%%%%%%%%%%%%%%%%%%%%%%

\def\theequation{\Alph{section}.\arabic{equation}}
\def\thesubsection{\Alph{section}.\arabic{subsection}}
\setcounter{section}{1}
\setcounter{equation}{0}
\vspace{2ex}

%%%%%%%%%%%%%%%%%%%%%%%%%%%%%%%%%%%%%%%

\subsection{The Hadamard finite part}

%%%%%%%%%%%%%%%%%%%%%%%%%%%%%%%%%%%%%%%

We have introduced in this paper some singular distributions denoted by
the symbol ${\cal P}\!f$. They are generated by the Hadamard finite
part of a divergent integral. See refs.\ \c{Schwartz,Zemanian} for more
details on these distributions. The idea is the following, suppose that
we have a function $h(x)$ which is Lebesgue integrable on all the
intervals $(a\+\epsilon,b)$, with $\epsilon\!>\!0$, but which is not
integrable on $(a,b)$ ($a$, $b$ are some real constants). Let us
consider the integral
\be
J(\epsilon)\equiv \int_{a+\epsilon}^{b} dx \: h(x),
\ee
which diverges in the limit $\epsilon \rightarrow 0^{+}$, and assume
that it is possible to separate it in two parts
\be
J(\epsilon)=I(\epsilon)+F(\epsilon),
\label{I+F}
\ee
where $I(\epsilon)$ is a finite linear combination of negative powers
of $\epsilon$ and positive powers of $\ln \epsilon$ and $F(\epsilon)$
has a finite well defined limit as $\epsilon \rightarrow 0^{+}$ (in the
sense that is independent of the way by which is obtained). Then the
Hadamard finite part of the divergent integral
$\int_{a}^{b} dx \: h(x)$ is defined by
\be
{\cal F}\!p \!\!\int_{a}^{b} dx \: h(x)\equiv \lim_{\epsilon
\rightarrow 0^{+}}
\! F(\epsilon)= \lim_{\epsilon \rightarrow 0^{+}}\! \left\{
\int_{a+\epsilon}^{b} dx \: h(x)- I(\epsilon) \right\},
\ee
that is, we throw away the divergent terms in (\ref{I+F}) and then
take the limit $\epsilon \rightarrow 0^{+}$. The definition is readily
extended to functions $h(x)$ defined on all the real axis which have
singular points. All we have to do is to decompose the integral over
$I\hspace{-0.7 ex} R$ as a sum of integrals of the kind considered
above. Then one can define distributions denoted by
${\cal P}\!f\left[h(x)\right]$ in the following way
\be
\int_{-\infty}^{\infty} dx \: {\cal P}\!f\left[h(x)\right] \,
\varphi(x)\equiv
{\cal F}\!p \!\!\int_{-\infty}^{\infty} dx \: h(x)\, \varphi(x),
\ee
where $\varphi(x)$ is an arbitrary tempered test function. The
definition can be easily generalized to the case of several variables
whenever the divergent integrals can be reduced to one-dimensional
divergent integrals.

As an example we calculate  the derivative of the distribution
\mbox{$(b^{2}\-1)^{-1/2}\, \t(b\-1)$} in detail. Notice that it is a
distribution in a
one-dimensional space (and also in a four-dimensional space if we set
$b\!\equiv\!t/r$) since the integral $\int_{-\infty}^{\infty} db
\:(b^{2}\-1)^{-1/2}\, \t(b\-1)\, \varphi(b)=\int_{1}^{\infty} db
\:(b^{2}\-1)^{-1/2}\, \varphi(b)$ is convergent. To start with, it is
easy to show that
\be
\frac{\!\!d}{db}\left[(b^{2}-1)^{-1/2}\, \t(b-1)\right]={\cal P}\!f
\left[
-b \, (b^{2}-1)^{-3/2} \, \t(b-1)\right].   \label{derivative}
\ee
For this, let us consider the following integral
\begin{eqnarray}
\hspace*{-4.2ex}\int_{1+\epsilon}^{\infty} db  \left[ -b \,
(b^{2}\-1)^{-3/2}\right]
\varphi(b)\!\!\!\!\!\!\!\!\!&&=\int_{1+\epsilon}^{\infty} db \:
\frac{\!\!d}{db}\left[ (b^{2}\-1)^{-1/2}\right]\varphi(b) \nonumber\\
\!\!\!\!\!\!\!\!\!&&= - (b^{2}\-1)^{-1/2}
\,\varphi(b){\Big\vert}_{b=1+\epsilon}\!\!-\!
\int_{1+\epsilon}^{\infty} db \:
(b^{2}\-1)^{-1/2}\,\varphi'(b),
\label{Pf}
\end{eqnarray}
where we have integrated by parts in the last step. This shows that
\begin{eqnarray}
\hspace*{-6ex}\int_{-\infty}^{\infty} db \: {\cal P}\!f \left[
-b\, (b^{2}\-1)^{-3/2} \, \t(b\-1)\right]
\varphi(b)\!\!\!\!\!\!\!\!\!&&\equiv {\cal F}\!p \!\!
\int_{1}^{\infty} db
\left[-b\, (b^{2}\-1)^{-3/2}\right]  \varphi(b) \nonumber\\
\!\!\!\!\!\!\!\!\!&&=-\int_{1}^{\infty} db \:
(b^{2}\-1)^{-1/2} \,\varphi'(b)  \nonumber\\
\!\!\!\!\!\!\!\!\!&&=\int_{-\infty}^{\infty} db \:
\frac{\!\!d}{db}\left[(b^{2}\-1)^{-1/2}\, \t(b\-1)\right]\varphi(b),
\end{eqnarray}
where the definition of the derivative of a distribution has been used,
and this proves (\ref{derivative}).
An explicit expression for the distribution
${\cal P}\!f \left[ -b \,(b^{2}\-1)^{-3/2} \, \t(b\-1)\right]$ can be
derived as a distributional limit. In fact, from (\ref{Pf}) we have
\begin{eqnarray}
&&\hspace*{-12ex}\int_{-\infty}^{\infty}db \: {\cal P}\!f \!\left[
-b\,(b^{2}\-1)^{-3/2} \, \t(b\-1)\right]
\varphi(b) \nonumber\\
&&=\lim_{\epsilon \rightarrow 0^{+}} \Biggl\{
\int_{1+\epsilon}^{\infty}  db \,\left[ -b\,(b^{2}\-1)^{-3/2}\right]
\varphi(b) +\varphi(b)\,(b^{2}\-1)^{-1/2}{\Big\vert}_{b=1+\epsilon}
\Biggr\} \nonumber\\
&&=\lim_{\epsilon \rightarrow 0^{+}} \left\{
\int_{1+\epsilon}^{\infty}  db \left[ -b\,(b^{2}\-1)^{-3/2} \right]
\varphi(b)+\frac{1}{\sqrt{2\epsilon}}\, \varphi(1) \right\} \nonumber\\
&&=\lim_{\epsilon \rightarrow 0^{+}}
\int_{-\infty}^{\infty}  db \Biggl[ -b\,(b^{2}\-1)^{-3/2}\,
\t(b\-1\-\epsilon)
+\frac{1}{\sqrt{2\epsilon}}\, \d(b\-1) \Biggr] \varphi(b),
\label{Pf calculation}
\end{eqnarray}
which gives the sought expression
\begin{eqnarray}
\hspace*{-5ex}\frac{\!\!d}{db}\left[(b^{2}-1)^{-1/2}\, \t(b-1)\right]
\!\!\!\!\!\!\!\!\!&&=
{\cal P}\!f \left[-b \,(b^{2}\-1)^{-3/2} \, \t(b-1)\right]  \nonumber\\
\!\!\!\!\!\!\!\!\!&&= \lim_{\epsilon \rightarrow 0^{+}}
\left[ -b \,(b^{2}\-1)^{-3/2} \,\t(b\-1\-\epsilon)+
\frac{1}{\sqrt{2\epsilon}}\, \d(b-1) \right],
\end{eqnarray}
where this distributional limit has to be understood in the sense of
(\ref{Pf calculation}).

In a similar way it is easy to show that
\begin{eqnarray}
\hspace*{-1ex}\frac{\!\!d^{2}}{db^{2}}\left[(b^{2}-1)^{-1/2}\, \t(b-1)
\right]\!\!\!\!\!\!\!\!\!&&=
{\cal P}\!f \left[ (2b^{2}+1)\,(b^{2}-1)^{-5/2} \, \t(b-1)\right]
\nonumber\\
\!\!\!\!\!\!\!\!\!&&= \lim_{\epsilon \rightarrow 0^{+}}
\Biggl[ (2b^{2}+1)\,(b^{2}-1)^{-5/2}\,\t(b\-1\-\epsilon) \nonumber\\
&&\hspace*{8.5ex}-\left(\frac{1}{(2\epsilon)^{3/2}}
+\frac{1}{8}\,\frac{1}{\sqrt{2\epsilon}}\right)\,\d(b-1)
+\frac{3}{2}\,\frac{1}{\sqrt{2\epsilon}}\, \d'(b-1) \Biggr].
\nonumber\\
\end{eqnarray}
Other Hadamard finite part distributions appearing in this paper are
given by the following distributional limits:
\begin{eqnarray}
&&\hspace*{-3.5ex}{\cal P}\!f \left[ (b^{2}\-1)^{-3/2}\,\t(b\-1)\right]
= \lim_{\epsilon \rightarrow 0^{+}}
\left[ (b^{2}\-1)^{-3/2}\,\t(b\-1\-\epsilon)-
\frac{1}{\sqrt{2\epsilon}}\, \d(b\-1) \right],   \nonumber\\
&&\hspace*{-3.5ex}{\cal P}\!f \left[\kappa\, p(b)\, (b^{2}\-1)^{-3/2}
\,\t(b\-1)\right]
= \kappa\, p(b)\,{\cal P}\!f \left[ (b^{2}\-1)^{-3/2}\,\t(b\-1)\right]
\nonumber\\
&&\hspace*{26.3ex}= \lim_{\epsilon \rightarrow 0^{+}}
\left[ \kappa\, p(b)\, (b^{2}\-1)^{-3/2}\,\t(b\-1\-\epsilon)\-
\kappa\,\frac{p(1)}{\sqrt{2\epsilon}}\, \d(b\-1) \right],   \nonumber\\
&&\hspace*{-3.5ex}g(b)\equiv {\cal P}\!f \left[(b^{2}\-1)^{-5/2}\,
\t(b\-1)\right]
\=\lim_{\epsilon \rightarrow 0^{+}}
\Biggl[  (b^{2}\-1)^{-5/2}\,\t(b\-1\-\epsilon)   \nonumber\\
&&\hspace*{30ex}\;-\left(\frac{1}{3}\,\frac{1}{(2\epsilon)^{3/2}}\-
\frac{5}{8}\,
\frac{1}{\sqrt{2\epsilon}}\right)\d(b\-1)
+\frac{1}{2}\,\frac{1}{\sqrt{2\epsilon}}\, \d'(b\-1)\Biggr],
\nonumber\\
&&\hspace*{-3.5ex}f_{C}(b)\!\equiv\!{\cal P}\!f
\left[b\,(b^{2}\-1)^{-5/2}\,(2b^{4}\-5b^{2}\+C/4)
      \,\t(b\-1)\right]        \nonumber\\
&&\hspace*{2.2ex}=\lim_{\epsilon \rightarrow 0^{+}}\Biggl\{
b \, (b^{2}\-1)^{-5/2}\,(2b^{4}\-5b^{2}\+C/4)\,\t(b\-1\-\epsilon)
\nonumber\\
&&\hspace*{6.4ex}+ \left[ \left(1\-\frac{C}{12}\right)
\frac{1}{(2\epsilon)^{3/2}}\+
   \frac{1}{8} \left(5\+\frac{C}{4}\right) \frac{1}{\sqrt{2\epsilon}}
\right]\!\d(b\-1)
    -\frac{3}{2}\left(1\-\frac{C}{12}\right)\frac{1}{\sqrt{2\epsilon}}
\,\d'(b\-1)
    \Biggr\},           \nonumber\\
\end{eqnarray}
where $\kappa\!\!\neq\!\!0$ is a constant and $p(b)$ is a polynomial
in $b$ with $p(1)\!\!\neq\!\!0$.

%%%%%%%%%%%%%%%%%%%%%%%%%%%%%%%%%%%%%%%%%%%%%%

\subsection{The propagator $H(x,\bar{\mu})$}

%%%%%%%%%%%%%%%%%%%%%%%%%%%%%%%%%%%%%%%%%%%%%%

Let us consider the expression
\be
\frac{1}{\pi}\, \t(x^{0})\, \frac{d}{d(x^{2})}\, \d(x^{2}) =
\frac{1}{\pi}\,\lim_{\lambda \rightarrow 0^{-}}
\frac{\!\!d}{d\lambda} \left[\t(x^{0})\, \d (x^{2}+\lambda) \right],
\ee
which is not a distribution in a four-dimensional space. To see
this we consider the effect of integrating an arbitrary tempered
test function $\phi(x)$ with such an expression,
\begin{eqnarray}
\int \! d^{4}x \:\frac{1}{\pi}\, \t(x^{0})\, \frac{d}{d(x^{2})}\,
\d(x^{2}) \: \phi(x)=\frac{1}{4 \pi} \!\int d^{2}\Omega
\int_{0}^{\infty} \frac{dr}{r}
\: \phi {\big\vert}_{t=r}-\frac{1}{4 \pi} \int \! d^{2}\Omega
\int_{0}^{\infty} \!\! dr
\: \frac{\,\p \phi}{\p t} {\Bigg\vert}_{t=r}, \nonumber\\
\mbox{}   \label{integral}
\end{eqnarray}
where $r\!\equiv \mid \!\vec{x}\!\mid$. The integral in the first term
is, in general, divergent. The Hadamard
finite part of the divergent integral $\int_{0}^{\infty} dr \, r^{-1}
\, \varphi(r)$, where $\varphi(r)$ is a tempered test function, is
\be
{\cal F}\!p \!\!\int_{0}^{\infty} \frac{dr}{r}\: \varphi(r)=
\lim_{\epsilon \rightarrow
0^{+}} \left[ \int_{\epsilon}^{\infty} \frac{dr}{r}\:
\varphi(r)+\ln \!\epsilon \;\varphi (0) \right].
\ee
Taking $\varphi(r)\!\equiv \! \int \! d^{2}\Omega \:\phi
{\big\vert}_{t=r}$, one can define the finite part of the integral
(\ref{integral}). In this way we can define a distribution, which we
call ${\cal P}\!f \!\left[\frac{1}{\pi}\, \t(x^{0})\,
\frac{d}{d(x^{2})}\, \d(x^{2}) \right]$, as
\begin{eqnarray}
&&\hspace*{-12ex}\int \! d^{4}x \:{\cal P}\!f \!\left[\frac{1}{\pi}\,
\t(x^{0})\,
\frac{d}{d(x^{2})}\, \d(x^{2}) \right]  \phi(x)   \nonumber\\
&&\equiv {\cal F}\!p \!\!\int \! d^{4}x \:\frac{1}{\pi}\, \t(x^{0})\,
\frac{d}{d(x^{2})}\,
\d(x^{2}) \: \phi(x)  \nonumber\\
&&=\frac{1}{4 \pi}\: {\cal F}\!p \!\!\int_{0}^{\infty} \frac{dr}{r}\:
\varphi(r) - \frac{1}{4 \pi} \int \! d^{2}\Omega
\int_{0}^{\infty} \!\! dr\: \frac{\,\p \phi}{\p t} {\Bigg\vert}_{t=r}
\nonumber\\
&&= \lim_{\epsilon \rightarrow 0^{+}} \left[
\int_{\mid \vec{x}\mid\geq\epsilon}
 d^{4}x \:\frac{1}{\pi}\, \t(x^{0})\, \frac{d}{d(x^{2})}\,
\d(x^{2}) \, \phi(x)+\ln \!\epsilon \,\phi (0)\right]   \nonumber\\
&&= \lim_{\epsilon \rightarrow 0^{+}} \int \! d^{4}x \Biggl[
\frac{1}{\pi}\, \t(x^{0})\,
\t(\mid \!\vec{x}\!\mid\!-\epsilon) \,\frac{d}{d(x^{2})}\,
\d(x^{2}) +\ln \!\epsilon \: \d^{4}(x) \Biggr]  \phi(x),
\end{eqnarray}
{}from which we see that
${\cal P}\!f \!\left[\frac{1}{\pi}\,\t(x^{0})\,
\frac{d}{d(x^{2})}\, \d(x^{2}) \right]\=\lim_{\epsilon
\rightarrow 0^{+}}
\!\Bigl[\frac{1}{\pi}\, \t(x^{0})\,
\t(\mid \!\vec{x}\!\mid\!-\epsilon) \,\frac{d}{d(x^{2})}\,
\d(x^{2}) +\ln \!\epsilon \: \d^{4}(x) \Bigr]$.
{}From eqs.\ (\ref{F2}) and (\ref{F3}) we see that this distribution
satisfies
\be
x_{\a} \, {\cal P}\!f \!\left[\frac{1}{\pi}\,\t(x^{0})\,
\frac{d}{d(x^{2})}\, \d(x^{2}) \right]=x_{\a} \,\frac{1}{\pi}\,
\t(x^{0})\, \frac{d}{d(x^{2})}\, \d(x^{2})= x_{\a}\,H(x,\bar{\mu}).
\label{equality}
\ee
Notice that (\ref{equality}) is an equality
between distributions: in the second member, the factor $x_{\a}$ kills
the divergences. This equality shows that the propagator
$H(x,\bar{\mu})$ must be equal to the distribution
${\cal P}\!f \!\left[\frac{1}{\pi}\,\t(x^{0})\,
\frac{d}{d(x^{2})}\, \d(x^{2}) \right]$ plus a distribution
$\tilde{H}(x,\bar{\mu})$ which satisfies $x_{\a}\,
\tilde{H}(x,\bar{\mu})\=0$.
It is easy to prove that such a distribution can only be a constant
times the delta distribution, so we have
\be
H(x,\bar{\mu})={\cal P}\!f \!\left[\frac{1}{\pi}\,\t(x^{0})\,
\frac{d}{d(x^{2})}\, \d(x^{2}) \right]+C(\bar{\mu})\, \d^{4}(x),
\label{propagator3}
\ee
where $C(\bar{\mu})$ is a constant involving the arbitrary mass scale
$\bar{\mu}$. Note that we could not have applied this argument to
equation (\ref{F3}) because $\frac{1}{\pi}\,\t(x^{0})\,
\frac{d}{d(x^{2})}\,\d(x^{2})$ is not a distribution. The constant
$C(\bar{\mu})$ can be computed by evaluation of the integral
$\int \! d^{4}x \: H(x,\bar{\mu})\, \t(R\,-\!\mid \!\vec{x}\!\mid)$,
with $R$ constant, using, on the one hand, expression
(\ref{propagator3}) and, on the other hand, the following Fourier
transform representation
\be
H(x,\bar{\mu})= -\frac{1}{2} \int \frac{d^{4}p}{(2\pi)^{4}}\: e^{-i p x}
\left[\ln\left(\frac{\mid p^{2}\!\mid}{\bar{\mu}^{2}}\right)-i \pi \,
\t(p^{2}) \:{\rm sign}(p^{0})\right].   \label{equiv H}
\ee
This can be easily shown to be equivalent to the definition of
$H(x,\bar{\mu})$ in eq.\ (\ref{H}) using
$\lim_{\epsilon \rightarrow 0^{+}}\ln \left[ -(p^{2}
+i\epsilon) \right]\= \ln \!\mid \!p^{2}\!\!\mid\!\-\,i \pi \,
\t(p^{2})$ and $2 \,\t(-p^{0})\-1\=-{\rm sign}(p^{0})$. We obtain
$C(\bar{\mu})\=\ln \bar{\mu}\+\gamma\-1$, so we finally have
\begin{eqnarray}
\hspace*{-4ex}H(x,\bar{\mu})&=&{\cal P}\!f \!\left[\frac{1}{\pi}\,
\t(x^{0})\,
\frac{d}{d(x^{2})}\, \d(x^{2}) \right]+
\left[\ln \!\bar{\mu}\+\gamma\-1 \right] \d^{4}(x),      \nonumber\\
&=&\lim_{\epsilon \rightarrow 0^{+}} \!\left\{\frac{1}{\pi}\,
\t(x^{0})\,
\t(\mid \!\vec{x}\!\mid\!-\epsilon) \,\frac{d}{d(x^{2})}\,
\d(x^{2}) +\left[\ln \!\bar{\mu}\epsilon\+\gamma\-1 \right] \d^{4}(x)
\right\}.
\label{propagator4}
\end{eqnarray}
This expression can be shown to be equivalent to
\be
H(x,\bar{\mu})=\lim_{\lambda \rightarrow 0^{-}} \!\left\{\frac{1}{\pi}
\,\t(x^{0})\,
\frac{\!\!d}{d\lambda} \d (x^{2}+\lambda)
+\left[\frac{1}{2}\,\ln \!\left(-\lambda \bar{\mu}^{2}\right)\+
\gamma\-1 \right]\d^{4}(x)
\right\},
\ee
which is, in fact, the definition for the propagator $H$ used by
Horowitz in ref.\ \c{Horowitz 80}.

%%%%%%%%%%%%%%%%%%%%%%%%%%%%%%%%%%%%%%%%%%%%%%%%%%%

\subsection{Calculation of the integral $I(x,y)$}

%%%%%%%%%%%%%%%%%%%%%%%%%%%%%%%%%%%%%%%%%%%%%%%%%%%

We show here the calculation of the integral
\be
I(x,y)\equiv \int d^{4}x'\: H(-x',\bar{\mu})\, \d(x+x') \d(y+y')
\ee
which appeared in section 3. One can calculate this integral in two
ways: using the representation (\ref{equiv H}) for the propagator $H$
as a Fourier transform or using expression (\ref{propagator4}).
With the representation (\ref{equiv H}) for $H$ we have
\be
I(x,y)= - \int \frac{d^{2}p}{(2\pi)^{2}} \: e^{i \vec{p}\cdot \vec{x} }
\, \ln \frac{\mid \vec{p}\mid}{\bar{\mu}},
\ee
where $\vec{p}\!\equiv\! (p^{x},p^{y})$, $\vec{x}\!\equiv\! (x,y)$ are
here vectors in a two-dimensional space. That is, $I$ is the Fourier
transform of a logarithm in a two-dimensional space. The result, which
can be found in ref.\ \c{Schwartz}, is
\begin{eqnarray}
I(x,y)&=&\frac{1}{2 \pi} \, {\cal P}\!f \left(\frac{1}{r^{2}} \right)+
\left(
\ln \frac{\bar{\mu}}{2} +\gamma \right) \d(x)\d(y) \nonumber\\
&=& \lim_{\epsilon \rightarrow 0^{+}} \left\{\frac{1}{2 \pi} \,
\frac{1}{r^{2}}
\, \t(r-\epsilon ) + \left( \ln \frac{\bar{\mu} \epsilon}{2} +\gamma
\right) \d(x)\d(y)
\right\},  \label{result}
\end{eqnarray}
where $r\!\equiv\! \sqrt{x^{2}+y^{2}}$.

For completeness let us check that we can get this result using the
representation (\ref{propagator4}) for H,
\be
I(x,y)=\lim_{\epsilon \rightarrow 0^{+}} \left\{ (\ln \bar{\mu}
\epsilon+\gamma \-1)\,
\d(x)\d(y)+\frac{1}{2\pi}\int_{0}^{\infty}\!\! dz \:
\frac{1}{(z^{2}\+r^{2})^{3/2}} \,
\t(r^{2}\+z^{2}\-\epsilon^{2}) \right\}.
\ee
Introducing $1\=\t(r\-\epsilon)\+\t(\epsilon\-r)$
in the second term one has
\begin{eqnarray}
I(x,y)\!\!\!\!\!\!\!\!\!&&=\!\!\lim_{\epsilon \rightarrow 0^{+}}\!
\Biggl\{ (\ln \bar{\mu} \epsilon\+\gamma \-1)\,
\d(x)\d(y)+\frac{1}{2\pi} \, \t(r\-\epsilon)\hspace{-1pt}
\int_{0}^{\infty}\!\! dz \:
\frac{1}{(z^{2}\+r^{2})^{3/2}}    \nonumber\\
&&\hspace{8ex}+(1\-\ln 2)\, {\cal G}(x,y;\epsilon)  \Biggr\}
\nonumber\\
\!\!\!\!\!\!\!\!\!&&=\!\!\lim_{\epsilon \rightarrow 0^{+}}\! \left\{
\frac{1}{2\pi}\,\frac{1}{r^{2}}\,\t(r\-\epsilon)+ (\ln \bar{\mu}
\epsilon\+\gamma \-1)\,\d(x)\d(y)+
(1\-\ln 2)\, {\cal G}(x,y;\epsilon) \right\}, \nonumber\\
\end{eqnarray}
where
\begin{eqnarray}
{\cal G}(x,y;\epsilon)&\equiv& \frac{1}{2\pi} \, \frac{1}{(1\-\ln 2)}
\,\t(\epsilon\-r)
\int_{0}^{\infty}\!\! dz \: \frac{1}{(z^{2}\+r^{2})^{3/2}}
\,\t(r^{2}\+z^{2}\-\epsilon^{2}) \nonumber\\
&=& \frac{1}{2\pi} \, \frac{1}{(1\-\ln 2)}\,\frac{1}{r^{2}}\,
\t(\epsilon\-r)\,
\left(1-\frac{\sqrt{\epsilon^{2}-r^{2}}}{\epsilon} \right).
\end{eqnarray}
Now to prove (\ref{result}) it remains to be seen that
$\lim_{\epsilon \rightarrow 0^{+}}{\cal G}(x,y;\epsilon)\=\d(x)\d(y)$.
For this we use the following theorem (see, for
example, ref.\ \c{Schwartz 2}). Suppose we have a function
$f(x;\epsilon)$, with $x \!\in\!I\hspace{-0.7 ex} R^{n}$, that
satisfies the following conditions:
1) $f(x;\epsilon)\!\geq\!0$ for $\| x \|\!\leq\!\kappa$, with
$\kappa\!>\!0$ some fixed constant;
2) in all the sets $a\!\leq\!\| x \|\!\leq\! \frac{1}{a}$,
being $a\!>\! 0$ an arbitrary finite constant, $f(x;\epsilon)$
converges uniformly to zero as $\epsilon \!\rightarrow\! 0^{+}$; and
\mbox{3) $\lim_{\epsilon \rightarrow 0^{+}}
\int_{\| x \|\leq a,\: a>\epsilon}  d^{n}x \: f(x;\epsilon)=1$.} Then
it can be shown that $\lim_{\epsilon \rightarrow 0^{+}}
f(x;\epsilon)\=\d^{n}(x)$. On the one hand we see that
${\cal G}(x,y;\epsilon)\!\geq\!0$ and that ${\cal G}(x,y;\epsilon)\=0$
for $r\!\geq\!a\!>\!\epsilon$, so, in the limit
$\epsilon \!\rightarrow\! 0^{+}$, ${\cal G}(x,y;\epsilon)$ converges
uniformly to zero for $r\!\geq\!a\!>\!0$. On the other hand it is easy
to prove that
$\int_{r\leq a,\: a>\epsilon} dx\,dy \:{\cal G}(x,y;\epsilon)\=1$,
thus we have that
$\lim_{\epsilon \rightarrow 0^{+}}{\cal G}(x,y;\epsilon)\=\d(x)\d(y)$.

\vspace{5ex}

%%%%%%%%%%%%%%%%%%%%%%%%%%%%%%%%%%%

{\noindent \Large \bf Appendix B}
\appendix

%%%%%%%%%%%%%%%%%%%%%%%%%%%%%%%%%%%

\setcounter{section}{2}
\setcounter{equation}{0}
\vspace{2ex}

\noindent Riemann tensor in the linear approximation for $h_{\mu \nu}$
with cylindrical symmetry.

\vspace{2ex}

It is easy to check that the Riemann components (\ref{Riemann}) for
$h_{\mu \nu}$ with cylindrical symmetry can be written in terms of
$h_{1}\!\equiv\!h^{t}_{t}$, $h_{2}\!\equiv\!h^{z}_{z}$,
$h_{3}\!\equiv\!h^{r}_{r}\+h^{\t}_{\t}$, $h_{4}\!\equiv\!h^{r}_{r}\-
h^{\t}_{\t}$ and $h_{5}\!\equiv\!h^{r}_{t}\=-h^{t}_{r}$
in the following form:
\begin{eqnarray}
&&\!\!\!\!\!\!\!\!\!\!R_{\,tztz}=\frac{1}{2} \, h_{2,tt}, \;\;\;\;\;\;\;
R_{\,xzxz}= \frac{1}{2} \left( \cos^{2}\! \t \; h_{2,rr}+ \sin^{2}\!
\t \; \frac{h_{2,r}}{r} \right),  \nonumber\\
&&\!\!\!\!\!\!\!\!\!\!R_{\,xzyz}= \frac{1}{4} \sin 2\t \left( h_{2,rr}
-\frac{h_{2,r}}{r}  \right), \nonumber\\
&&\!\!\!\!\!\!\!\!\!\! R_{\,txtx}= \frac{1}{4} \left( h_{3,tt}\+
\cos 2\t \; h_{4,tt}\-2 \cos^{2}\! \t \; h_{1,rr}\-
              2 \sin^{2}\! \t \; \frac{h_{1,r}}{r} \-4 \cos^{2}\! \t
\; h_{5,tr}\- 4 \sin^{2}\! \t \; \frac{h_{5,t}}{r}
              \right),\nonumber\\
&&\!\!\!\!\!\!\!\!\!\!R_{\,txty}=\frac{1}{4} \sin 2\t \left( h_{4,tt}
-h_{1,rr}
             +\frac{h_{1,r}}{r}-2 \, h_{5,tr}
             +2 \, \frac{h_{5,t}}{r} \right),  \;\;\;\;\;\;\;
R_{\,tzxz}= \frac{1}{2} \cos \t \; h_{2,tr},  \nonumber\\
&&\!\!\!\!\!\!\!\!\!\!R_{\,txxy}= \frac{1}{4} \sin \t \left( -h_{3,tr}
+h_{4,tr}
             +2 \, \frac{h_{4,t}}{r} \right), \;\;\;\;
R_{\,xyxy}= \frac{1}{4} \left( h_{3,rr}+\frac{h_{3,r}}{r} -h_{4,rr}-3 \,
             \frac{h_{4,r}}{r} \right)\hspace{-1.6pt}.
\nonumber
\end{eqnarray}
$R_{\,yzyz}$, $R_{\,tyty}$, $R_{\,tzyz}$ and $R_{\,tyyx}$ can be
obtained from the previous expressions interchanging $\cos \t$ by
$\sin \t$ in $R_{\,xzxz}$, $R_{\,txtx}$, $R_{\,tzxz}$ and $R_{\,txxy}$
respectively.

%%%%%%%%%%%%%%%%%%%%%%%%%%%%%%%%%%%

\end{document}